\documentclass[letterpaper,english,a4paper, reprint, aip, jcp, amsmath, amssymb]{revtex4-1}
\usepackage[T1]{fontenc}
\usepackage[latin9]{inputenc}
\usepackage{color}
\usepackage{babel}
\usepackage{float}
\usepackage{amsmath}
\usepackage{amssymb}
\usepackage{graphicx}
\usepackage[unicode=true]
 {hyperref}

\makeatletter

\pdfpageheight\paperheight
\pdfpagewidth\paperwidth

\newcommand{\noun}[1]{\textsc{#1}}

\pdfoutput=1

\usepackage{dcolumn}% Align table columns on decimal point
\usepackage{bm}% bold math
\usepackage{mathptmx}

\usepackage{xcolor}
\hypersetup{colorlinks,
linkcolor={red!50!black},
citecolor={blue!50!black},
urlcolor={blue!50!black}}

\makeatother

\begin{document}

\title{Identification of kinetic order parameters for non-equilibrium dynamics}

\author{Fabian Paul}

\address{FU Berlin, Department of Mathematics and Computer Science, Arnimallee
6, 14195 Berlin, Germany}

\address{University of Chicago, 929 East 57th Street Chicago, IL 60637, USA}

\author{Hao Wu}

\address{Tongji University, School of Mathematical Sciences, Shanghai, 200092,
P.R. China}

\address{FU Berlin, Department of Mathematics and Computer Science, Arnimallee
6, 14195 Berlin, Germany}

\author{Maximilian Vossel}

\address{Max Planck Institute for Biophysical Chemistry, Am Fassberg 11 D-37077
Göttingen, Germany}

\author{Bert L. de Groot}

\address{Max Planck Institute for Biophysical Chemistry, Am Fassberg 11 D-37077
Göttingen, Germany}

\author{Frank Noé$^{*}$}

\address{FU Berlin, Department of Mathematics and Computer Science, Arnimallee
6, 14195 Berlin, Germany}

\address{Rice University, Department of Chemistry, Houston, Texas 77005, USA}
\email{frank.noe@fu-berlin.de}

\begin{abstract}
A popular approach to analyze the dynamics of high-dimensional many-body
systems, such as macromolecules, is to project the trajectories onto
a space of slowly-varying collective variables, where subsequent analyses
are made, such as clustering or estimation of free energy profiles
or Markov state models (MSMs). However, existing ``dynamical'' dimension
reduction methods, such as the time-lagged independent component analysis
(TICA) are only valid if the dynamics obeys detailed balance (microscopic
reversibility) and typically require long, equilibrated simulation
trajectories. Here we develop a dimension reduction method for non-equilibrium
dynamics based on the recently developed Variational Approach for
Markov Processes (VAMP) by Wu and Noé. VAMP is illustrated by obtaining
a low-dimensional description of a single file ion diffusion model
and by identifying long-lived states from molecular dynamics simulations
of the KcsA channel protein in an external electrochemical potential.
This analysis provides detailed insights into the coupling of conformational
dynamics, the configuration of the selectivity filter, and the conductance
of the channel. We recommend VAMP as a replacement for the less general
TICA method.
\end{abstract}
\maketitle

\section{Introduction}

Much understanding about molecular kinetics has been gained by modeling
kinetics with Markov state models (MSMs),\citep{Schuette:JComputPhys:99,Singhal:JChemPhys:04,NoeHorenkeSchutteSmith_JCP07_Metastability,Prinz:JChemPhys:11}
rate equation models \citep{Buchete:JPhysChemB:08} or diffusion map-based
models.\citep{RohrdanzClementi_JCP134_DiffMaps,PretoClementi_PCCP14_AdaptiveSampling,ZhengEtAl_JPCB11}
A key element in all of these methods is that the dynamics are modeled
in a low-dimensional space of collective variables.\citep{RohrdanzEtAl_AnnRevPhysChem13_MountainPasses,Peters:JChemPhys:06,NoeClementi_COSB17_SlowCVs}
In MSMs and rate equation models, there is a direct link between
the kinetic model, consisting of a set of states and transition probabilities,
and the underlying microscopic dynamical equations \emph{via} the
spectral decomposition of Markov operators.\citep{Schuette:ZIB:99}
As a result of this theory, the natural collective variables to describe
the long-time dynamics are the eigenfunctions of the Markov operator.\citep{NoeClementi_COSB17_SlowCVs}
In practice, the eigenfunctions and eigenvalues of the (Markov) transfer
operator can be approximately computed directly from molecular dynamics
(MD) simulation data by means of the Noé-Nüske variational approach.\citep{Noe:MultiscaleModelSim:13,Nueske:JChemTheoryComput:14}
This has led to wide application of spectral methods in the molecular
dynamics community, in particular the time-lagged independent component
analysis (TICA).\citep{Molgedey:PhysRevLett:94,PerezHernandez:JChemPhys:13,Schwantes:JChemTheoryComput:13} 

However, application of TICA is only truly justified if the dynamics
fulfill the principle of detailed balance (microscopic reversibility)
and if the dynamical equations are stationary, i.e. do not change
as a function of time. Moreover, since TICA is a data-driven method,
the reversibility and stationarity must be approximately met in the
(finite) simulation data. Ignoring this limitation can result in systematic
errors. For instance, if TICA is applied to non-equilibrium data (such
as data that consist of short trajectories that were not initialized
from the equilibrium distribution) the computed eigenvalues and eigenfunctions
incur large biases.\citep{Wu:JChemPhys:17} A valid alternative method
that can be applied to non-equilibrium data is Koopman reweighting.\citep{Wu:JChemPhys:17}
This method removes estimation bias, but empirically induces a large
variance as seen in the results of Ref.\citep{Wu:JChemPhys:17}.

The restriction to equilibrium data impedes the analysis of interesting
and biologically relevant molecular systems whose function relies
on non-reversible dynamics. Ion conduction in channel proteins is
an example for such a process, since the ion current is driven by
external and perhaps time-dependent electric fields and chemical potentials.
Therefore, a dynamical dimension reduction method that is similar
to TICA, but is directly applicable to non-equilibrium dynamics or
even to non-equilibrium data would be desirable.

Recently, a new approach for dimensionality reduction of dynamic systems
was proposed by Wu and Noé.\citep{Wu:ArXiv:17} The \emph{variational
approach to Markov processes} (VAMP) dispenses with the assumptions
of stationarity and reversibility. This was made possible by reformulating
the problem of dimensionality reduction as a regression problem. Similarly
to the reversible methods like TICA, VAMP can be directly applied
to MD simulation data; it is hence a data-driven method. Wu et al\emph{.}
showed that there exists an optimal low-rank approximation to the
solution of the above regression problem. This gives rise to a low-dimensional
space of order parameters that are chosen such that the regression
error is minimized. Mathematically this space can be found by performing
a restricted singular value decomposition \citep{DeMoor:Stanford:89,Abdi:GSVD:07}
of a regression matrix learned from the simulation data. 

Mardt et al\emph{. }\citep{Mardt:NatCommun:18} showed that VAMP can
be used to train a deep neural network to find informative order parameters
and derive a coarse-grained MSM for the conformational dynamics of
the alanine dipeptide and the folding of the N-terminal domain of
ribosomal protein L9 (NTL9). In their work, Mardt et al\emph{.} focused
on demonstrating that VAMP can be successfully used to select highly
non-linear transformations to approximate the singular functions.
However the MD simulations that they used were reversible and stationary.

In this work, we show that VAMP works as a dimension reduction method
for non-equilibrium data that may or may not originate from an equilibrium
system. The goal is to establish an alternative to TICA which can
be applied to reduce the dimension of the data and keep the slow processes,
no matter whether the data are too short to be equilibrated, or if
the underlying process is fundamentally out of equilibrium. We also
extend the Chapman-Kolmogorov test, which is frequently used to validated
MSMs,\citep{Prinz:JChemPhys:11} to validate the Markov property of
the dimensionality-reduced model obtained with VAMP. We demonstrate
VAMP by identifying the slow collective variables for two non-equilibrium
systems: 1) the asymmetric simple exclusion process (ASEP) which is
a simple model of single file diffusion. 2) non-equilibrium MD simulation
data \citep{Koepfer:Science:2014} of the KcsA potassium ion channel
in which an ion current is driven through the channel pore. Furthermore,
using a simple model of diffusion in a low-dimensional energy landscape,
we compare the biases of VAMP and TICA when applied to a ensemble
of short trajectories that were initiated from an non-equilibrium
distribution.

\section{Theory}

We first lay a theoretical framework with the most important mathematical
results. The more practically inclined reader is advised to skip to
the Methods section. The theoretical framework is formulated in the
language of dynamical operators. The advantage of this formulation
is that theoretical properties can be obtained by using linear methods
- albeit in infinite many dimensions. The main theoretical result
is a variational principle, which can then be used for the formulation
of linear or nonlinear solvers (such as VAMPnets \citep{Mardt:NatCommun:18}).
In order to go into even more theoretical detail, please refer to
Ref.\citep{Wu:ArXiv:17}.

\subsection{Exact dynamics in full configuration space}

Let $\mathbf{x}$ be the coordinates in which the MD algorithm is
Markovian (atom positions, velocities, box coordinates etc.) Let $p(\mathbf{x},t)$
be the probability density of finding the system in state $\mathbf{x}$
at time $t$. We are interested in $p(\mathbf{x},\tau n)$, the density
at times $\tau n$ that are integer multiples of some lag time $\tau$.
At these times, the time evolution of $p$ can be described with the
following integral equation:
\begin{align}
p(\mathbf{x}^{\prime},t+\tau) & =\int p(\mathbf{x}^{\prime}\mid\mathbf{x})p(\mathbf{x},t)\,\mathrm{d}x=\mathcal{P}_{\tau}[p]
\end{align}
Here $\mathcal{P}_{\tau}$ stands for the propagation operator (or
propagator) which can be thought of as the discrete-time analog of
the Fokker-Planck operator. $p(\mathbf{x}^{\prime}\mid\mathbf{x})$
denotes the conditional probability density of visiting an infinitesimal
phase space volume around point $\mathbf{x}^{\prime}$ at time $t+\tau$
given that the phase space point $\mathbf{x}$ was visited at the
earlier time $t$.

An equivalent description of the time evolution is given by the following
integral equation which defines the Koopman operator $\mathcal{K}_{\tau}$:
\begin{align}
g(\mathbf{x},t+\tau) & =\int p(\mathbf{x}^{\prime}\mid\mathbf{x})f(\mathbf{x}^{\prime})\,\mathrm{d}x^{\prime}=\mathcal{K}_{\tau}[f]
\end{align}
$f$ is an observable, i.e. in general a function of positions and
momenta. The result $g(\mathbf{x},t+\tau)$ can be interpreted as
the expectation value of $f$ at time $t+\tau$ computed from an ensemble
that was propagated for a time $\tau$ after having been started at
time $t$ from the single point $\mathbf{x}$:
\begin{align}
\mathcal{K}_{\tau}[f](\mathbf{x}) & =\mathbb{E}_{t+\tau}\left(f\mid p(\mathbf{x},t)=\delta(\mathbf{x})\right)
\end{align}
Here $\delta$ is the (vectorial) Dirac delta function.

Both the propagator and the Koopman operator fulfill the Chapman\textendash Kolmogorov
equation 
\begin{align}
\mathcal{P}_{\tau_{1}+\tau_{2}} & =\mathcal{P}_{\tau_{1}}\mathcal{P}_{\tau_{2}}\\
\mathcal{K}_{\tau_{1}+\tau_{2}} & =\mathcal{K}_{\tau_{1}}\mathcal{K}_{\tau_{2}}\label{eq:Koopman-MarkovProperty}
\end{align}
For stationary dynamics, this implies that expectations of any observable
$f$ can be computed for all times from the Koopman operator.
\begin{equation}
g(\mathbf{x},n\tau)=\mathbb{E}_{n\tau}\left(f\mid p(\mathbf{x},0)=\delta(\mathbf{x})\right)=\mathcal{K}_{\tau}^{n}f
\end{equation}
Expectations for an ensemble, that was started from an arbitrary probability
density $p_{0}$, can be computed from the following scalar product:
\begin{equation}
\mathbb{E}_{n\tau}\left(f\mid p(\mathbf{x},0)=p_{0}(\mathbf{x})\right)=\int p_{0}(\mathbf{x})g(\mathbf{x},n\tau)\,\mathrm{d}x
\end{equation}
For the computation of instantaneous and time-lagged variances and
covariances,\citep{Noe:PNAS:11} similar equations that use the Koopman
operator can be derived.

\subsection{Formulation of ensemble propagation as a regression problem}

Can a dynamic model be built using only expectation values that were
computed from simulation data? This question has been addressed in
a series of papers preceding our VAMP theory that have developed the
so-called Koopman analysis.\citep{Mezic_NonlinDyn05_Koopman,TuEtAl_JCD14_ExactDMD}
We seek a small matrix $\mathbf{K}_{\tau}\in\mathbb{R}^{k\times k}$,
called the Koopman matrix, that fulfills the equation
\begin{align}
\mathcal{K}_{\tau}\mathbf{g} & \approx\mathbf{K}_{\tau}^{\top}\mathbf{f}\label{eq:DimRed-MarkovProperty}
\end{align}
in a sense that we explain below. Here $\mathbf{f}$ and $\mathbf{g}$
are vectors of observables, that can be arbitrary functions of the
conformation. $\mathbf{f}(\mathbf{x})=(f_{1}(\mathbf{x}),f_{2}(\mathbf{x}),\ldots)^{\top}$
and similarly for $\mathbf{g}$. We use the shorthand notation $\left(\mathcal{K}_{t}\mathbf{g}\right)_{i}:=\mathcal{K}_{t}g_{i}$
which means that the Koopman operator is applied element-wise to $\mathbf{g}$.

More formally, for fixed $\mathbf{f}$ and $\mathbf{g}$, the optimal
data-dependent matrix $\mathbf{K}_{\tau}$ can be computed by minimizing
the following error
\begin{equation}
\epsilon=\mathbb{E}_{\rho_{0}}\left[\left\Vert \mathcal{K}_{\tau}\mathbf{g}-\mathbf{K}_{\tau}^{\top}\mathbf{f}\right\Vert ^{2}\right].\label{eq:Koopman-error}
\end{equation}
$\rho_{0}$ is the empirical distribution of the simulation data,
excluding time steps $t_{i}>T-\tau$ where $T$ is the length of the
(single) time series. By inserting the definitions of $\rho_{0}$
and $\mathcal{K}_{\tau}$ into \eqref{eq:Koopman-error}, one finds
that
\begin{equation}
\epsilon=\sum_{{t\text{ s.t. }\atop 0\leq t\leq T-\tau}}\left\Vert \mathbf{g}\left(\mathbf{x}(t+\tau)\right)-\mathbf{K}_{\tau}^{\top}\mathbf{f}\left(\mathbf{x}(t)\right)\right\Vert ^{2}\label{eq:Koopman-error-data}
\end{equation}
which shows that the error is purely data-dependent. 

Eqs. \ref{eq:Koopman-error} and \ref{eq:Koopman-error-data} have
the form of a regression problem: a future window from the time series
is regressed against the current window of the time series. This formulation
avoids any assumption of microscopic reversibility. 

\subsection{Optimal low-dimensional observables}

Unlike the Koopman matrix, the observable functions $\mathbf{f}$
and $\mathbf{g}$ cannot be chosen by only minimizing the regression
error defined by \eqref{eq:Koopman-error}, because the minimal $\epsilon=0$
can be trivially obtained by an uninformative model with $\mathbf{f}(\mathbf{x})\equiv\mathbf{g}(\mathbf{x})\equiv1$.

In Ref.\citep{Wu:ArXiv:17}, the error of the approximate Koopman
operator provided by model \eqref{eq:DimRed-MarkovProperty} was analyzed.
It was shown that the model with the smallest approximation error
in Hilbert-Schmidt norm is given by $\mathbf{f}=\boldsymbol{\psi}=(\psi_{1},\ldots,\psi_{k})^{\top}$,
$\mathbf{g}=\boldsymbol{\phi}=(\phi_{1},\ldots,\phi_{k})^{\top}$
and $\mathbf{K}_{\tau}=\mathrm{diag}(\boldsymbol{\sigma})=\mathrm{diag}(\sigma_{1},\ldots,\sigma_{k})$
for a given $k$, and the corresponding approximation of the Koopman
operator is
\begin{equation}
\mathcal{K}_{\tau}g\approx\sum_{i=1}^{k}\sigma_{i}\langle g,\phi_{i}\rangle_{\rho_{1}}\psi_{i},\label{eq:K-SVDd}
\end{equation}
where $\sigma_{i}$ is the $i$'th largest singular value of $\mathcal{K}_{\tau}$,
$\psi_{i}$ and $\phi_{i}$ are the corresponding left and right singular
functions respectively. (The singular value decomposition is to be
understood of being applied after a whitening transformation of $\mathbf{f}$
and $\mathbf{g}$. See method subsection \ref{subsec:TCCA} for details.)
$\rho_{1}$ is the empirical distribution of simulation data excluding
time steps $t_{i}<\tau$, and $\langle f,g\rangle_{\rho_{1}}=\int f(\mathbf{x})g(\mathbf{x})\rho_{1}(\mathbf{x})\,\mathrm{d}\mathbf{x}$.

It can be shown that the largest singular value $\sigma_{1}$ is always
1 and that the corresponding left and right singular functions are
constant and identical to 1 for all $\mathbf{x}$.\citep{Wu:JChemPhys:17}
Only the singular components $\sigma_{i}$, $\psi_{i}$, $\phi_{i}$
with $i>1$ contain kinetic information.

If $\boldsymbol{\psi}$ and $\boldsymbol{\phi}$ are approximated
with a finite linear combination of ansatz functions, a corresponding
finite-dimensional singular value decomposition of the whitened Koopman
matrix can be used to compute the optimal superposition coefficients
(see subsection \ref{subsec:TCCA} for details).

\subsection{The kinetic map induced by the singular functions}

For a Markov process, we can measure the difference between two points
$\mathbf{x}$ and $\mathbf{y}$ by the kinetic distance \citep{Noe:JChemTheoryComput:15}
$D_{\tau}(\mathbf{x},\mathbf{y})$ where
\begin{equation}
D_{\tau}^{2}(\mathbf{x},\mathbf{y})=\int\frac{\left(p(\mathbf{z}|\mathbf{x})-p(\mathbf{z}|\mathbf{y})\right)^{2}}{\rho_{1}(\mathbf{z})}\,\mathrm{d}\mathbf{z}.\label{eq:kinetic-dist}
\end{equation}
$D_{\tau}(\mathbf{x},\mathbf{y})=0$ means $\mathbf{x}$ and $\mathbf{y}$
are equivalent for predicting the future evolution of the process.
By using the singular components of $\mathcal{K}_{\tau}$, the square
of the kinetic distance can be written as
\begin{equation}
D_{\tau}^{2}(\mathbf{x},\mathbf{y})=\sum_{i}\sigma_{i}^{2}\left(\psi_{i}(\mathbf{x})-\psi_{i}(\mathbf{y})\right)^{2}\label{eq:kinetic-map}
\end{equation}
(see Appendix \ref{sec:Proof-of-kinetic-map} for proof). If all but
the $k$ leading singular values are close $0$, we have
\begin{equation}
D_{\tau}^{2}(\mathbf{x},\mathbf{y})\approx\left\Vert \mathrm{diag}(\boldsymbol{\sigma})\boldsymbol{\psi}(\mathbf{x})-\mathrm{diag}(\boldsymbol{\sigma})\boldsymbol{\psi}(\mathbf{y})\right\Vert ^{2},\label{eq:kinetic-map-low-dim}
\end{equation}
where $\mathrm{diag}(\boldsymbol{\sigma})$ denotes the $k\times k$
diagonal matrix with the singular values on its diagonal. That means,
all the points $\mathbf{x}$ can be embedded into a $k$-dimensional
Euclidean space by the kinetic map $\mathbf{x}\to\mathrm{diag}(\boldsymbol{\sigma})\boldsymbol{\psi}(\mathbf{x})$
with the structure of the kinetic distance preserved. Note that the
extension of Ref.\citep{Noe:JChemTheoryComput:15} to the commute
distance \citep{NoeClementi_JCTC16_KineticMap2} is not directly applicable
to VAMP because the commute distance relies on the computation of
relaxation timescales, which relies on the eigenvalue decomposition
of the Markov operator and cannot be directly done with the singular
value decomposition.

Also note that the kinetic distance defined in Eq. \ref{eq:kinetic-dist}
depends on the empirical distribution of the data $\rho_{1}$. Therefore,
$D_{\tau}^{2}(\mathbf{x},\mathbf{y})$ in general depends on how the
system dynamics were sampled. For systems that possess an unique stationary
distribution (see for example the ASEP model in subsection \ref{subsec:ASEP}),
$\rho_{1}$ can be set to the stationary distribution to define a
kinetic distance that is independent from the sampling.

Furthermore, it is worth noting that the coherent sets of non-reversible
Markov processes can also be identified from the $k$ dominant singular
components, and more details can be seen in Ref.\citep{Koltai:Computation:18}.
Also, note that the right singular functions $\boldsymbol{\phi}$
induce a kinetic map with respect to time-reversed propagation of
the dynamics (unlike the kinetic map induced by $\boldsymbol{\psi}$
that uses conventional forward-time propagation).

\section{Methods}

In this work we use VAMP as a method for computing optimal kinetic
order parameters for non-equilibrium dynamics using a linear combination
of input features. However, in general, the scope of VAMP is larger:
order parameters are not restricted to be linear combinations but
can also be formed from a non-linear combination of features as was
demonstrated by Mardt et al\emph{. }\citep{Mardt:NatCommun:18} by
training a deep neural network (VAMPnet) to capture the conformational
dynamics of the alanine dipeptide and the N-terminal domain of ribosomal
protein L9 (NTL9). Another application of VAMP is the scoring of input
features (see publication ``Variational Selection of Features for
Molecular Kinetics'' by Scherer et al\emph{.} in this issue). 

Using VAMP to find kinetic order parameters from a linear combination
of molecular features is also called time-lagged canonical covariance
analysis (TCCA) \citep{Mardt:NatCommun:18} and works as follows.

\subsection{Dimension reduction using the Variational Approach for Markov Processes
(VAMP)\label{subsec:TCCA}}

Let $\boldsymbol{\chi}(t)$ be a multivariate time series where every
element $\chi_{i}(t)$ is the time series of one molecular feature.
Features can be Cartesian or internal coordinates (such as distances
or dihedral angles) of the molecular system or functions thereof (such
as the sine and cosine of dihedral angles or a step function that
converts a distance into a contact). From the input features $\boldsymbol{\chi}(t)$,
first the means $\boldsymbol{\mu}_{0}$ and $\boldsymbol{\mu}_{1}$
are computed from all data excluding the last and first $\tau$ steps
of every trajectory, respectively:

\begin{align}
\boldsymbol{\mu}_{0} & :=\frac{1}{T-\tau}\sum_{t=0}^{T-\tau}\boldsymbol{\chi}(t)\\
\boldsymbol{\mu}_{1} & :=\frac{1}{T-\tau}\sum_{t=\tau}^{T}\boldsymbol{\chi}(t)
\end{align}
Next, the instantaneous covariance matrices $\mathbf{C}_{00}$ and
$\mathbf{C}_{11}$ and the time-lagged covariance matrix $\mathbf{C}_{01}$
are computed as follows:

\begin{align}
\mathbf{C}_{00} & :=\frac{1}{T-\tau}\sum_{t=0}^{T-\tau}\left[\boldsymbol{\chi}(t)-\boldsymbol{\mu}_{0}\right]\left[\boldsymbol{\chi}(t)-\boldsymbol{\mu}_{0}\right]^{\top}\label{eq:C00}\\
\mathbf{C}_{11} & :=\frac{1}{T-\tau}\sum_{t=\tau}^{T}\left[\boldsymbol{\chi}(t)-\boldsymbol{\mu}_{1}\right]\left[\boldsymbol{\chi}(t)-\boldsymbol{\mu}_{1}\right]^{\top}\label{eq:C11}\\
\mathbf{C}_{01} & :=\frac{1}{T-\tau}\sum_{t=0}^{T-\tau}\left[\boldsymbol{\chi}(t)-\boldsymbol{\mu}_{0}\right]\left[\boldsymbol{\chi}(t+\tau)-\boldsymbol{\mu}_{1}\right]^{\top}\label{eq:C01}
\end{align}
After that, a Koopman matrix $\bar{\mathbf{K}}$ is computed in the
basis of whitened \citep{Karhunen:Whitening:01,Wu:ArXiv:17} input
features

\begin{equation}
\bar{\mathbf{K}}:=\mathbf{C}_{00}^{-\frac{1}{2}}\mathbf{C}_{01}\mathbf{C}_{11}^{-\frac{1}{2}}\label{eq:Kbar-def}
\end{equation}
Then, the singular value decomposition (SVD) of $\bar{\mathbf{K}}$
is performed, giving orthonormal matrices $\mathbf{U}^{\prime}$ and
$\mathbf{V}^{\prime}$ as well as $\mathbf{S}=\mathrm{diag}(\boldsymbol{\sigma})$
such that

\begin{equation}
\bar{\mathbf{K}}=\mathbf{U}^{\prime}\mathbf{S}\mathbf{V}^{\prime}\label{eq:SVD}
\end{equation}
Finally, the input conformations are mapped to the left singular functions
$\boldsymbol{\psi}$ and right singular functions $\boldsymbol{\phi}$
as follows:

\begin{equation}
\boldsymbol{\psi}(t):=\mathbf{U}^{\prime\top}\mathbf{C}_{00}^{-\frac{1}{2}}\left[\boldsymbol{\chi}(t)-\boldsymbol{\mu}_{0}\right]\label{eq:psi-def}
\end{equation}

\begin{equation}
\boldsymbol{\phi}(t):=\mathbf{V}^{\prime\top}\mathbf{C}_{11}^{-\frac{1}{2}}\left[\boldsymbol{\chi}(t)-\boldsymbol{\mu}_{1}\right]\label{eq:phi-def}
\end{equation}
$\boldsymbol{\psi}(t)$ and $\boldsymbol{\phi}(t)$ are the sought-after
kinetic order parameters. Since the left singular functions $\boldsymbol{\psi}(t)$
induce a kinetic map for the (conventional) forward-time propagator,
they are the natural choice of order parameters if one wants to perform
a clustering of space to obtain state definitions. For simplicity,
we will call them VAMP components.

Note that the algorithm above performs a Canonical Correlation Analysis
(CCA) \citep{KnappEtAl_PsychBull78_CCA} in time, and is hence also
called Time-lagged CCA (TCCA) \citep{Wu:ArXiv:17}. The singular value
decomposition in the whitened basis \eqref{eq:Kbar-def}, \eqref{eq:SVD}
is also called the generalized \citep{Abdi:GSVD:07} or restricted
\citep{DeMoor:Stanford:89} SVD of $\mathbf{C}_{01}$ under constraints
imposed by $\mathbf{C}_{00}$ and $\mathbf{C}_{11}$.

\subsection{The variational score}

In the previous subsection \ref{subsec:TCCA}, VAMP was used to linearly
combine molecular features to compute kinetic order parameters. A
question that remained unanswered is how to select the best molecular
features to use as input. This question can be answered by computing
the variational score of the dimensionality-reduced kinetic model.
The VAMP-$r$ score is defined as the sum of the leading $m$ largest
singular values that have been taken to the power of $r$ (see Ref.\citep{Wu:ArXiv:17}
and the publication ``Variational Selection of Features for Molecular
Kinetics'' by Scherer et al\emph{.} in this issue).
\begin{equation}
\text{VAMP}_{r,\text{train}}=\sum_{i=1}^{m}\sigma_{i}^{r}
\end{equation}

In a situation with infinite sampling, where the singular values are
known without statistical error, the best selection of molecular features
is the one that maximizes the VAMP-$r$ score. In a practical setting
however where the time series data is finite, direct maximization
of the VAMP-$r$ score is not possible due to model overfitting.\citep{McGibbon:JChemPhys:15}
That is why the VAMP-$r$ score needs to be computed in a cross-validated
manner. 

Cross-validation works by splitting the trajectory data into two sets:
the training set, from which a dimensionality-reduced model is estimated,
and the test set, against which the model is tested. From the training
set, the matrices $\mathbf{U}^{\mathrm{train}}=\mathbf{C}_{00}^{-\frac{1}{2}}\mathbf{U}^{\prime}$
and $\mathbf{V}^{\mathrm{train}}=\mathbf{C}_{11}^{-\frac{1}{2}}\mathbf{V}^{\prime}$
are computed, where $\mathbf{C}_{00}$, $\mathbf{C}_{11}$, $\mathbf{U}^{\prime}$,
and $\mathbf{V}^{\prime}$ are computed from the training data according
to Eqs. \ref{eq:C00}, \ref{eq:C11}, and \ref{eq:SVD}. Next, the
test score is computed from the equation
\begin{equation}
\text{VAMP}_{r,\text{test}}=\sum_{i}\varkappa_{i}^{r}
\end{equation}
where $\varkappa_{i}$ is the $i$'th singular value of the matrix
product $\mathbf{A}\mathbf{B}\mathbf{C}$\textbf{}\footnote{This sum is by definition the $r$'th power of the so-called $r$-Schatten
norm \unexpanded{$\left\Vert .\right\Vert _{r}^{r}$}.} with 
\begin{align}
\mathbf{A} & =(\mathbf{U}^{\mathrm{train}\top}\mathbf{C}_{00}^{\mathrm{test}}\mathbf{U}^{\mathrm{train}})^{-\frac{1}{2}}\\
\mathbf{B} & =\mathbf{U}^{\mathrm{train}\top}\mathbf{C}_{01}^{\mathrm{test}}\mathbf{V}^{\mathrm{train}}\\
\mathbf{C} & =(\mathbf{V}^{\mathrm{train}\top}\mathbf{C}_{11}^{\mathrm{test}}\mathbf{V}^{\mathrm{train}})^{-\frac{1}{2}}
\end{align}
and where $\mathbf{C}_{00}^{\mathrm{test}}$, $\mathbf{C}_{01}^{\mathrm{test}}$,
and $\mathbf{C}_{11}^{\mathrm{test}}$ have been computed from the
test data \emph{via} Eqs. \ref{eq:C00}, \ref{eq:C01}, and \ref{eq:C11}
(with the caveat that the means $\boldsymbol{\mu}_{0}$, $\boldsymbol{\mu}_{1}$
of the \emph{training} data have to be subtracted). Finally, the $k$-fold
cross-validated test score is computed by repeating the splitting
of the data into test and training data $k$ times, computing one
test score for each partition of the data and then taking the average
of the individual test scores.

\subsection{The non-equilibrium Chapman-Kolmogorov test\label{subsec:CK-test}}

For stationary (but possibly non-reversible) dynamics the full-state-space
Koopman operator fulfills the Markov property \eqref{eq:Koopman-MarkovProperty}.
It shares this property with the propagator and with the transition
matrix of MSMs. 

In the context of molecular dynamics simulation, the Markov property
is often exploited to calculate long-time-scale properties ($\mathcal{K}_{n\tau}$)
from short-lag-time estimates ($\mathcal{K}_{\tau}$). One of the
most important long-time-scale properties is the stationary distribution
that can be computed from a MSM by applying the transition matrix
an infinite number of times to an initial probability distribution.

To extrapolate to higher multiples of the lag-time, the Markov property
needs to hold. While this property is guaranteed for the full-state-space
dynamical operators, it is not necessarily fulfilled for dimensionality-reduced
dynamical models like the transition matrix of a MSM or an approximated
Koopman operator. Therefore, the Markov property is typically tested
by comparing $\mathcal{K}_{n\tau}$ to $\mathcal{K}_{\tau}^{n}$ for
the multiples of the lag time $n\tau$.

The standard way to perform this test is to compare the direct estimate
of a time-lagged covariance 
\begin{equation}
\mathrm{cov}_{\mathrm{est}}(f,g;\,n\tau)=\langle f,\mathcal{K}_{n\tau}g\rangle_{\rho}\label{eq:cov-est}
\end{equation}
from the simulation data to the model-prediction of the same covariance
\begin{equation}
\mathrm{cov}_{\mathrm{pred}}(f,g;\,n\tau)=\langle f,\mathcal{K}_{\tau}^{n}g\rangle_{\rho}.\label{eq:cov-pred}
\end{equation}
$f$ and $g$ are some functions of the configuration-space coordinates.
When $f$ and $g$ are indicator function, this test is known under
the name Chapman-Kolmogorov test.\citep{Noe:PNAS:09} Here we propose
to perform the same comparison for the data-driven estimate of the
dimensionality-reduced Koopman operator. See appendix \ref{subsec:CK-test-implementation}
for details.

To make the test independent on the subjective choice of the functions
$f$ and $g$, the left and right singular functions of the Koopman
operator estimated at the lowest multiple of the lag time $1\times\tau$,
can be used as $f$ and $g$ respectively. This choice is in the spirit
of the Chapman-Kolmogorov test as it is typically applied to Markov
models of metastable molecular kinetics. There, the test is typically
applied to the probability of staying in one of the metastable states,
which constitutes a particular hard test that requires data that thoroughly
samples exit and entry events into the metastable states.\citep{Prinz:JChemPhys:11,Noe:PNAS:09}
Characteristic (indicator) functions of the metastable states are
related by a linear transform to the eigenfunctions of the transfer
operator.\citep{Schuette:ZIB:99} By analogy, we assume here that
using the singular functions in the Chapman-Kolmogorov test also constitutes
a particular hard test.

\subsection{Interpretation of the VAMP components and spectral clustering\label{subsec:clustering}}

For reversible dynamics, the theory of conformation dynamics describes
how the leading eigenfunctions can be used to understand which structural
changes are associated to the slowest processes and to find the metastable
states \emph{via} spectral clustering.\citep{Schuette:ZIB:99,Deuflhard:LinearAlgebraAppl:05}

For non-equilibrium dynamics, we can replace the eigenfunctions by
the left singular functions found by VAMP. As for TICA \citep{PerezHernandez:JChemPhys:13},
we can interpret the $i$th kinetic order parameter in terms of structural
changes by computing its correlation with all features $\chi_{j}$:
\begin{align}
\mathrm{corr}(\psi_{i},\chi_{j}) & =\frac{\frac{1}{T-\tau}\sum_{0\leq t<T-\tau}\psi_{i}(t)\left(\chi_{j}(t)-\bar{\chi}_{j}\right)}{\sqrt{\frac{1}{T-\tau}\sum_{0\leq t<T-\tau}\left(\chi_{j}(t)-\bar{\chi}_{j}\right)^{2}}}\\
 & =\frac{(\mathbf{C}_{00}^{\frac{1}{2}}\mathbf{U}^{\prime})_{ji}}{\sqrt{(\mathbf{C}_{00})_{jj}}}
\end{align}
and by visualizing the most-correlated features. In the last equation
$\bar{\chi}_{j}$ denotes the empirical mean of feature $\chi_{j}$
computed from the data in time steps $0\leq t_{i}<T-\tau$.

Furthermore, we can compute the long-lived states of non-equilibrium
dynamics by performing spectral clustering in the VAMP components,
in a similar way as it is done with the dominant eigenspace for equilibrium
dynamics in Ref.\citep{Weber:ZIB:02}. Let $\boldsymbol{\psi}$ be
the vector that contains the $n_{\mathrm{spec}}$ leading singular
functions (with singular values close to one, including the constant
singular function). Let $\mathbf{A}\in\mathbb{R}^{n_{\mathrm{spec}}\times n_{\mathrm{spec}}}$.
Then the vector of macro-state memberships $\mathbf{m}\in\mathbb{R}^{n_{\mathrm{spec}}}$
is given by
\begin{equation}
\mathbf{m}(t)=\mathbf{A}\boldsymbol{\psi}(t).
\end{equation}
See appendix \ref{subsec:inner-simplex} or Ref.\citep{Weber:ZIB:02}
for the algorithm to compute $\mathbf{A}$. The element $m_{i}(t)$
encodes the degree of membership of the conformation sampled at time
$t$ in the macro-state $i$. The memberships at every time step always
sum to one (which expresses the necessity of belonging to some macro-state
with certainty) and, depending on the specific algorithm that was
used to compute $\mathbf{A}$, are confined between 0 and 1 \citep{Deuflhard:LinearAlgebraAppl:05}
or not \citep{Weber:ZIB:02}. The memberships define the macro-states
in a fuzzy manner; that is every conformation belongs to macro-state
$i$ with a degree of membership given by $m_{i}(t)$. Fuzzy states
can be converted into crisp states by imposing a cutoff on the memberships
and treating conformations with memberships larger than the cutoff
as being part of the crisp state. Structural differences between states
can, e.g., be found using significant distance analysis.\citep{Stolzenberg:Bioinformatics:18}

\section{Results}

\subsection{Model of single file diffusion: the asymmetric simple exclusion process\label{subsec:ASEP}}

The asymmetric simple exclusion process (ASEP) is a generic model
for single file diffusion. It was originally formulated by MacDonald
et al\emph{.}\citep{MacDonald:Biopolymers:68} as a model for the
kinetics of protein synthesis and was independently introduced by
Spitzer\citep{Spitzer:AdvMath:70} in the mathematical literature.
Since then it has been extensively analyzed and applied to model phenomena
such as macromolecular transport, conductivity, traffic flow, sequence
alignment, and molecular motors (see Refs.\citep{Golinelli:JPhysA:06,KolomeiskyEtAl_ASEP_JPA98}
and references therein). 

\begin{figure}[!h]
\begin{centering}
\includegraphics{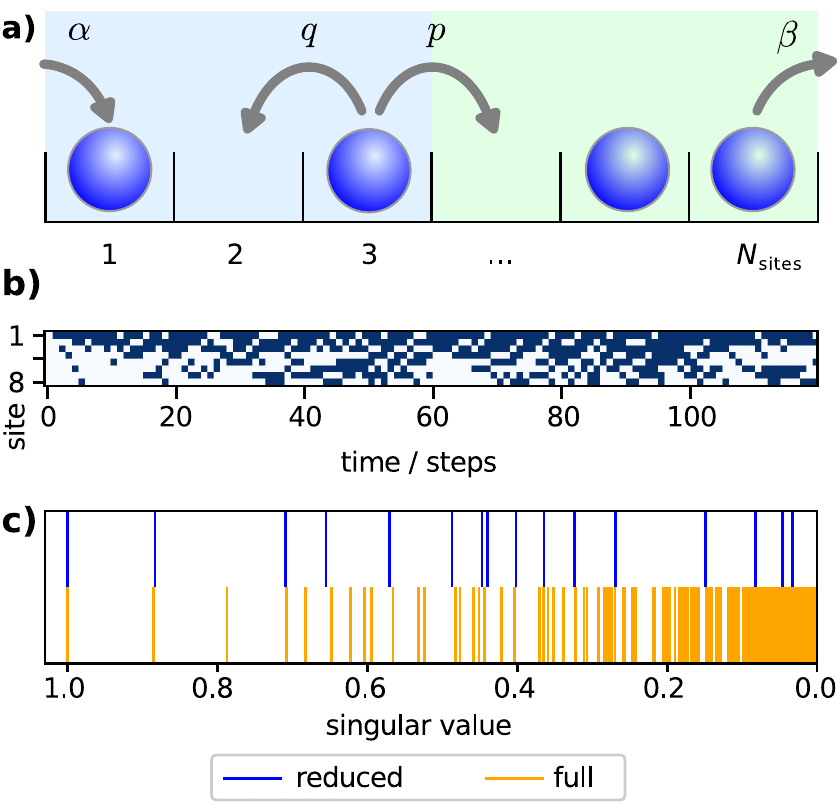}
\par\end{centering}
\caption{\label{fig:ASEP}a) The asymmetric exclusion process is a model for
single file diffusion. It consists of a linear chain of $N_{\mathrm{sites}}$
sites along which particles can move. Particles are inserted at position
1 with rate $\alpha$ and annihilated at site $N_{\mathrm{sites}}$
with rate $\beta$. b) the first 120 time steps of an exemplary realization
of a occupancy time trace for the ASEP with parameters $N_{\mathrm{sites}}=8$,
\textcolor{blue}{$\alpha=\beta=p=1$}, and \textcolor{blue}{$q=1/3$}.
c) singular values of the $17\times17$ Koopman model trained on the
time series (blue, upper spectrum) and singular values of the full
ASEP model (orange, lower spectrum).}
\end{figure}

The ASEP consists of a linear chain of $N_{\mathrm{sites}}$ sites
each of which can either be empty or occupied by exactly one particle,
resulting in a large state space with $2^{N_{\mathrm{sites}}}$ elements
(Fig.~\ref{fig:ASEP}a). If the first site is empty, a particle is
inserted with a rate $\alpha$. Particles can move to adjacent unoccupied
sites with rate $p$ in the forward and rate $q$ in the backward
direction. In the last site, particles are annihilated with rate $\beta$.
Hence, the ASEP is a driven (non-reversible) Markovian multi-particle
system. Here we show that VAMP can be used to train a low-dimensional
model that allows to reproduce the time-lagged covariances and auto-covariances
for a large range of lag-times.

We use the master equation formulation of the ASEP as our true reference
(Appendix \ref{sec:ASEP-model}). The model parameters are chosen
as $N_{\mathrm{sites}}=8$, $\alpha=\beta=p=1$, and $q=1/3$. Since
VAMP works with a finite lag-time, we convert the master equation
model to a transition matrix by taking the matrix-exponential of the
master equation coefficient matrix. From the transition matrix we
generate a long trajectory with $N_{\mathrm{steps}}=10^{6}$ steps.
The trajectory is encoded as a matrix of shape $N_{\mathrm{steps}}\times N_{\mathrm{sites}}$
where every row represents the occupancy pattern at a given time point
(see Fig.~\ref{fig:ASEP}b for an example of a transposed trajectory
matrix). 

We estimate an empirical Koopman matrix using VAMP at a lag time of
$\tau=1\,\mathrm{steps}$ and using a basis consisting of two groups
of features. The first group consists of the site occupancy vectors
(the columns of the matrix shown in Fig.~\ref{fig:ASEP}b). The second
set of features is an 9 dimensional vector that contains the ``one-hot''
encoded number of occupied sites. That is, element $i$ in the second
feature set is 1 if and only if there are $i$ occupied sites. Our
selection of features already constitutes a dimensionality-reduction,
since we estimate the Koopman model in the 8+9-dimensional space of
feature vectors and not in the $2^{8}$-dimensional state space. As
a consequence, the spectrum of the empirical Koopman model consists
of only 17 singular values. Not all singular values of the true model
can be reproduced (see Fig.~\ref{fig:ASEP}c), still, large singular
values approximately agree. The singular values of the empirical model
decay quickly with increasing rank (see top part of Fig.~\ref{fig:ASEP}c).
Therefore we discard the very small singular components and further
reduce the rank of the model to 11 dimensions (not counting the first
singular function which is the constant function). Despite these
two dimension reductions, we will show later that physically interesting
observables are correctly captured by the 11-dimensional model.

\begin{figure}
\begin{raggedright}
\includegraphics{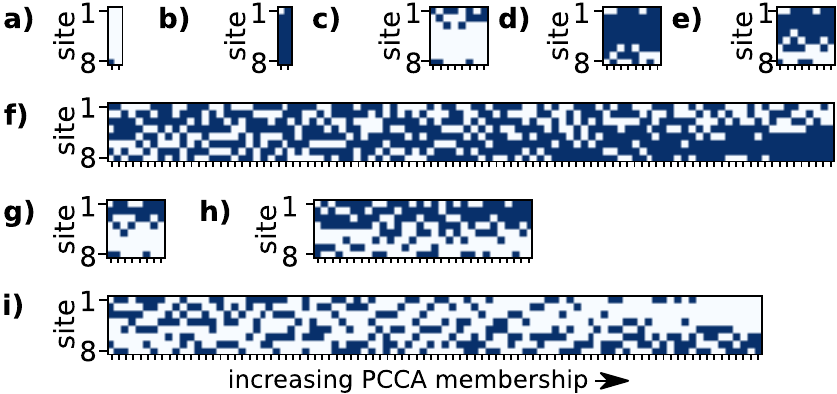}
\par\end{raggedright}
\caption{\label{fig:ASEP-states}The nine dominant long-lived macro-states
of the ASEP with parameters $N_{\mathrm{sites}}=8$, $\alpha=\beta=p=1$
and $q=1/3$. In each subfigure (a-i), one macro-state is shown. Occupancy
vectors of all microstates in a macro-state are ordered along the
micro-state axis (x-axis) with increasing memberships. Dark blue squares
mark occupied sites, white squares mark empty sites. The microstates
with the highest macro-state memberships (right-most patterns) are
characterized by long uninterrupted segments with the same occupancy
and a single jump from occupied to unoccupied (shock) for states (c)
to (i). Macro-states differ in the location of the shock.}
\end{figure}

To gain some physical understanding of the true singular functions
of the ASEP model, we cluster the space of the leading 9 VAMP components
of the true transition matrix with the PCCA+ algorithm without using
any further approximation (see KcsA application below for more details
on PCCA+ clustering). This allows to group all the possible site occupancy
patterns (micro-states) into 9 macro-states. We select 9 states because
in the true spectrum, a relatively large gap follows a denser cluster
of singular values at position 9 (see lower part of Fig.~\ref{fig:ASEP}c).
Macro-states are shown in Fig.~\ref{fig:ASEP-states}, with the micro-states
ordered from low macro-state membership to high macro-state membership
(from left to right). The top-membership micro-states are characterized
by long uninterrupted segments with the same occupancy (long occupied
/ long empty segments) and show only one alternation from occupied
to unoccupied (shock) along the queue. Macro-states differ in the
position of the shock. Micro-states with lower memberships resemble
the top membership states but show a noisier shock profile with more
alternations between occupied and unoccupied. Macro-states also differ
in the average number of occupied sites.

Next, we test whether our choice of the 17-dimensional basis that
consists of the occupancy vector and the one-hot encoded occupancy
affects the capability of PCCA+ to find the correct macro-states.
Therefore we repeat the PCCA+ clustering using the singular functions
that were approximated with the Koopman model. Since the simple-basis
does not allow to capture all leading 9 singular components correctly,
we perform the comparison in the space of the leading 3 components
(counting the constant component). Results are shown in Suppl. Fig.~5
and show good agreement between macro-states computed from the true
and the approximate model. With an increased number of macro-states,
the results deviate.

\begin{figure}[!h]
\begin{centering}
\includegraphics{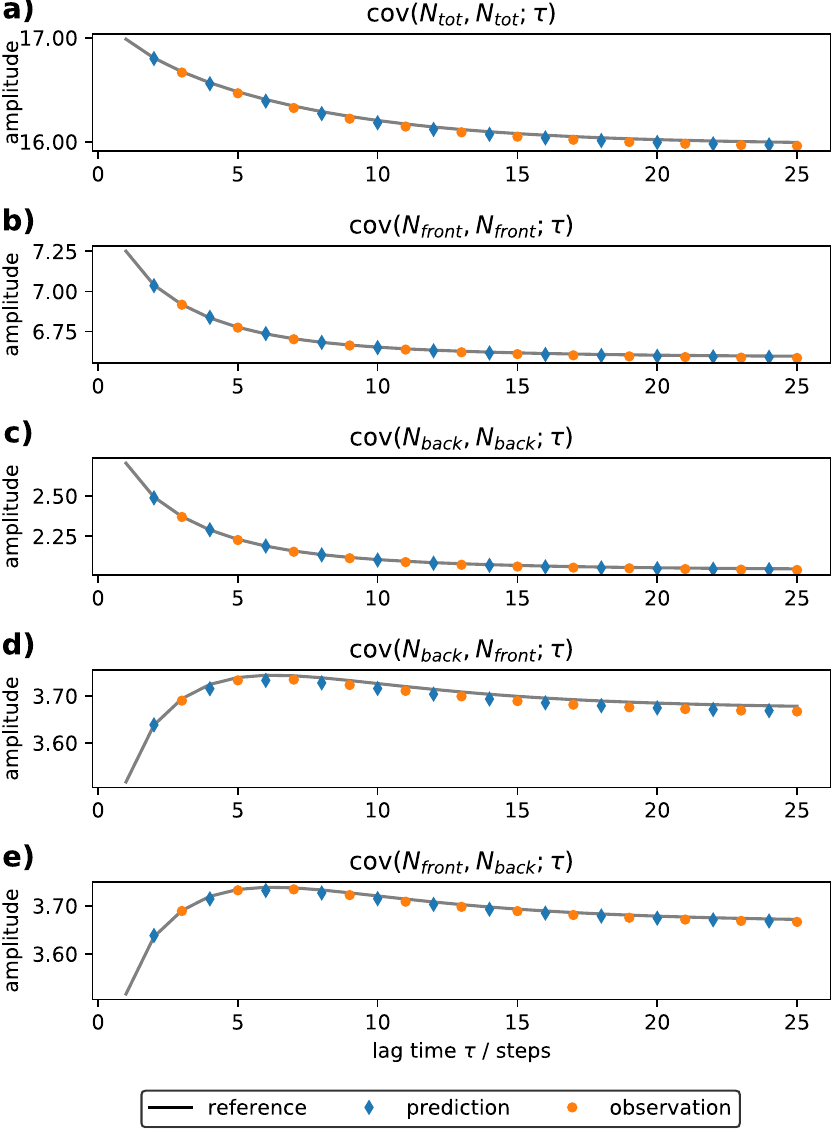}
\par\end{centering}
\caption{\label{fig:ASEP-CKtest}Chapman-Kolmogorov test results for the low-dimensional
Koopman matrix estimated from the ASEP model with parameters $N=8$,
$\alpha=\beta=p=1$, and $q=1/3$. $N_{\mathrm{front}}$ is the total
particle count in the first half of the chain (blue shaded area in
Fig.~\ref{fig:ASEP}a). $N_{\mathrm{back}}$ is the total particle
count in the second half (green shaded area in Fig.~\ref{fig:ASEP}a).
The true reference is computed from the full ASEP model, by using
Eq. \ref{eq:cov-pred} with the true Koopman operator.}
\end{figure}

To test the predictive power of the reduced model, we compare observed
time-lagged covariances to the model prediction of the same covariance
using the Chapman-Kolmogorov test. We pick one of the observables
$f=N_{\mathrm{front}}$ to be the number of particles in the first
half of the queue and the second observable $g=N_{\mathrm{back}}$
the number particles in the second half. Estimates for the observed
and the predicted time-lagged covariance of $N_{\mathrm{front}}$
and $N_{\mathrm{back}}$ computed from Eqs. \ref{eq:cov-est} and
\ref{eq:cov-pred} for multiple lag times are shown in Fig.~\ref{fig:ASEP-CKtest}b-e.
For comparison we also show the true covariances computed from the
full ASEP model without using the VAMP approximation (shown in gray
in Fig.~\ref{fig:ASEP-CKtest}). The Chapman-Kolmogorov test shows
that predictions from the dimensionality-reduced VAMP model agree
with the observed covariances computed from the time series data as
well as with the results from the full model.

\begin{figure}[!h]
\begin{centering}
\includegraphics{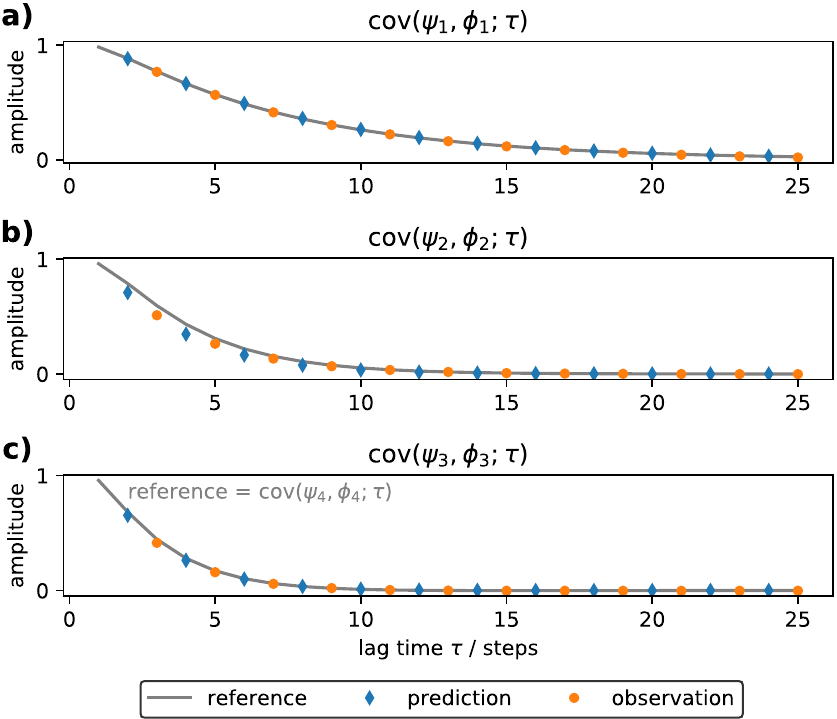}
\par\end{centering}
\caption{\label{fig:ASEP-CKtest-sing}Same as Fig.~\ref{fig:ASEP-CKtest}
but with singular functions chosen as observables. $\psi_{i}$ and
$\phi_{i}$ are left and right singular functions respectively of
the Koopman matrix estimated at the smallest lag time $\tau=1$. In
(c) $\mathrm{cov}(\psi_{4},\phi_{4};\tau)$ of the true ASEP transition
matrix serves as the reference, since the third singular function
of the dimensionality-reduced VAMP model predominantly matches the
fourth true singular function (see Suppl. Fig~1).}
\end{figure}

To make the Chapman-Kolmogorov test less dependent on the subjective
choice of observables $f$ and $g$, we repeat the test but this time
selecting the observables to be identical to the singular functions
$\psi_{i}^{(1)}$ and $\phi_{i}^{(1)}$, respectively, that were estimated
from the dimensionality-reduced model estimated at lag time $\tau=1\,\mathrm{steps}$.
That is, we compare $\mathrm{cov}_{\mathrm{est}}(\psi_{i}^{(1)},\phi_{i}^{(1)};\,n\tau)$
to $\mathrm{cov}_{\mathrm{pred}}(\psi_{i}^{(1)},\phi_{i}^{(1)};\,n\tau)$.
Results are shown in Fig.~\ref{fig:ASEP-CKtest-sing}. The figure
shows that the Chapman-Kolmogorov test succeeds for the all pairs
of singular functions, that is model predictions of covariances are
consistent with the re-estimated covariances for all lag times. Predictions
from the VAMP model are in good agreement with the true covariances
that were computed from the full ASEP model. The dimensionality-reduced
model does not correctly reproduce the second and third true singular
function but reproduces the fourth true singular function (see Suppl.
Fig.~1). To obtain this approximate agreement of the leading singular
functions, it was necessary to include the one-hot-encoded count of
occupied sites into the set of input features to VAMP. The mismatch
between the remaining singular functions and singular values of the
true and reduced model (see Fig.~\ref{fig:ASEP}c and Suppl. Fig.~1)
is a consequence of the very simple set of input features that was
used to estimate $\mathbf{K}_{\tau}$. Had the reduced model not been
trained on the 17-dimensional occupancy vectors but on the $2^{8}$-dimensional
full state space, the agreement would have been exact. Also using
a more expressive set of basis functions \citep{Mardt:NatCommun:18,Nueske:JChemPhys:16,Schwantes:JChemTheoryComput:15}
could be have produced a richer reduced model that captures more singular
components of the full model. Despite the simple approach, some observables
can be modeled correctly.

\begin{figure}
\begin{centering}
\includegraphics{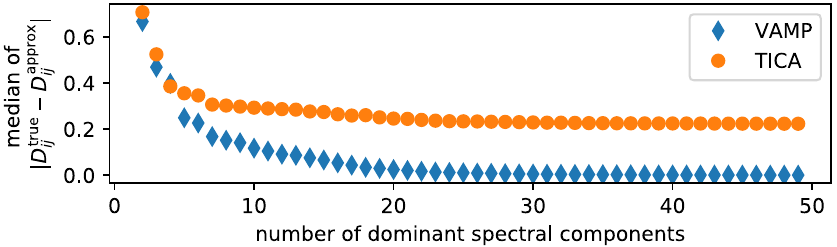}
\par\end{centering}
\caption{\label{fig:kinetic-dist-VAMP-vs-TICA}Comparison of the median difference
between the exact kinetic distance computed with Eq. \ref{eq:kinetic-dist}
to its low-rank VAMP approximation  and to its low-rank TICA approximation
as a function on the retained number of spectral components. $\rho_{1}$
in Eq. \ref{eq:kinetic-dist} was set to the stationary distribution
of the ASEP model.}
\end{figure}

Besides the estimation of a Koopman model, the typical use of VAMP
will be to compute kinetic order parameters for non-reversible kinetics.
To assess the improvement of these order parameters over the independent
components obtained from TICA, we compare the kinetic distance obtained
from TICA and VAMP to the true reference. We compute the true reference
of the kinetic distance by applying Eq. \ref{eq:kinetic-map-low-dim}
to the true ASEP transition matrix in a complete basis. We set $\rho_{1}$
in Eq. \ref{eq:kinetic-map-low-dim} to the true stationary distribution.
We compare this reference to the VAMP estimate computed from Eq. \ref{eq:kinetic-dist}
using the same full basis as well as to the TICA estimate. The TICA
estimate of the kinetic distance is computed from a modified Eq. \ref{eq:kinetic-dist}
with the singular values replaced by the TICA eigenvalues and the
right singular functions replaced by the TICA eigenfunctions. That
version is the default in the PyEMMA software.\citep{Scherer:JChemTheoryComput:15}
Results are shown in Fig.~\ref{fig:kinetic-dist-VAMP-vs-TICA}. As
implied by VAMP theory, the VAMP estimate converges to the true reference
as the number of singular components is increased. In contrast to
that, the TICA estimate does not converge to the true reference. This
is expected, since for non-equilibrium dynamics the kinetic distance
cannot be expressed using only the right eigenfunctions alone that
TICA provides.\citep{Noe:JChemTheoryComput:15} Full kinetic distances
between all states are given in Suppl. Fig.~4.

In summary, the application of VAMP to the ASEP model shows that VAMP
can accurately capture the dominant singular functions and can be
used to accurately compute time-lagged auto-covariances and cross-covariances
of physical quantities like the occupancy of the first and second
half of the queue. The ASEP is a genuinely non-reversible model. Therefore
its dimension-reduction can only be accomplished with methods like
VAMP that are capable of modeling non-reversible processes and do
not rely on detailed balance. A decomposition of the state space into
9 macro-states shows that the location of the shock (jump from occupied
to unoccupied segments) allows to approximately distinguish the macro-states
for the ASEP parameter settings that we chose. 

\subsection{Application of VAMP to a reversible system in the limit of non-equilibrium
sampling}

While the ASEP system is intrinsically non-equilibrium as its dynamical
equations violate detailed balance, we now investigate the performance
of VAMP when the underlying dynamics obey detailed balance, but the
data does not reflect the equilibrium distribution. In cases where
the metastable states are reversibly connected, reweighting methods
\citep{Wu:JChemPhys:17,Nueske:JChemPhys:17} and reversible maximum-likelihood
MSMs \citep{Bowman:JChemPhys:09,Prinz:JChemPhys:11} have been shown
to provide unbiased estimates and to recover the equilibrium kinetics
from non-equilibrium data. When transitions between states have only
been sampled in one direction, the current MSM practice is simply
to discard the not reversibly connected states.\citep{Bowman:JChemPhys:09,Prinz:JChemPhys:11}
In VAMP, this is not necessary because VAMP does not require a stationary
distribution to be computed.

\begin{figure}[H]
\begin{centering}
\includegraphics{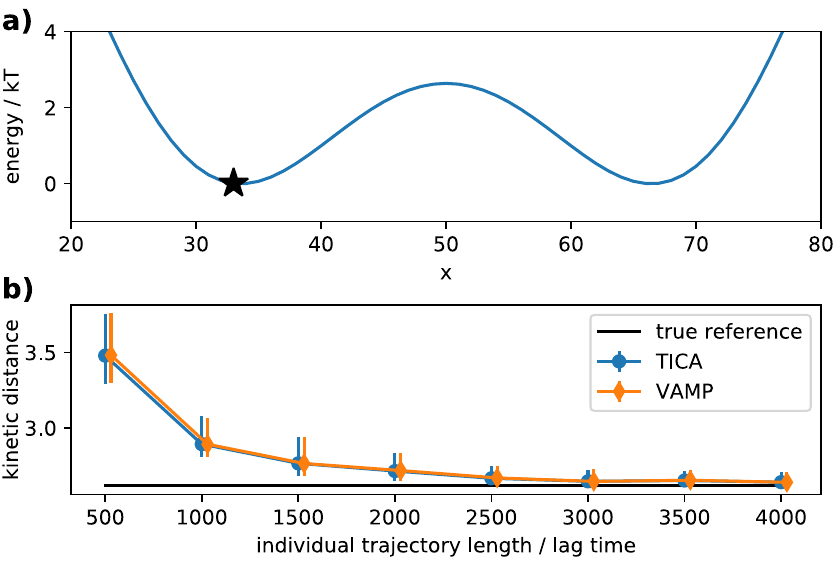}
\par\end{centering}
\caption{\label{fig:bad-sampling}a) Double-well energy landscape (parameters
from PyEMMA \citep{Scherer:JChemTheoryComput:15}) used to test VAMP
in the limit of non-equilibrium sampling. Trajectories were all started
from the minimum of the left well (star). b) Kinetic distance between
the two local minima of the energy landscape depending on the trajectory
length used for its estimation with VAMP or TICA. Plot markers mark
the median. Tips of the error bars mark the 10th and the 90th percentile.}
\end{figure}

Here we study the performance of VAMP on non-equilibrium data generated
from a 1-D double-well energy landscape (Fig.~\ref{fig:bad-sampling}a).
Trajectories were generated from the transition matrix which is provided
in the PyEMMA example datasets/models package \citep{Scherer:JChemTheoryComput:15}
using a lag time of $\tau=6$ steps. To produce non-equilibrium sampling,
we start all trajectories from the left well. The trajectory lengths
are $500\,\tau$ to $4000\,\tau$, which is on the order of the mean-first-passage
time to the right well. For each trajectory length, the aggregate
data over all trajectories is 90000 $\tau$. Each run is repeated
100 times to compute means and uncertainties.

We compare VAMP with TICA in terms of the kinetic distance between
the two energy wells. The kinetic distance is one of the few quantities
that can be computed from both VAMP and TICA, whereas eigenvalues
and singular values cannot directly be compared.

All estimates converge to the true reference value, as the trajectory
length is increased, when the sampling becomes increasingly representative
of the equilibrium kinetics (Fig.~\ref{fig:bad-sampling}). The true
reference is computed with Eq. \ref{eq:kinetic-dist} and with $\rho_{1}$
set to the true stationary distribution of the model transition matrix.
For non-equilibrium data (short trajectories), neither TICA nor VAMP
reproduce the equilibrium kinetic distance. In VAMP this is due to
the weighting of points with respect to the empirical distributions
$\rho_{1}$ which is in general different from the stationary distribution.
Strikingly, the results from VAMP and TICA are almost identical, both
in terms of the medians and of their statistical errors. This example
indicates no particular advantage of using VAMP over using TICA but
also no disadvantage.

Both VAMP and TICA can handle completely disconnected datasets (if
transitions in both directions between a pair of states are missing).
Every disconnected set leads to an additional singular value / eigenvalue
of value 1 (or close to one, due to projection errors). However, the
strength of VAMP lies elsewhere - in the analysis of inherently non-equilibrium
systems such as driven ion motion as exemplified by the ASEP model.

\subsection{Conformational changes of the KcsA potassium ion channel}

Ion channels are pore-forming transmembrane proteins that enable ions
to cross biomembranes. Ion channels are found both in the outer cell
membrane and in the membranes of the cell organelles. They are important
for functions such as cellular signaling, the regulation of osmotic
activity, and the propagation of action potentials in nerves and muscle
cells.\citep{Alberts:TheCell:08:11}

The first potassium channel protein to be crystallized is the bacterial
channel KcsA.\citep{Doyle:Science:98} The structure can be subdivided
into three consecutive parts: following the pore from the extracellular
to the intracellular side, one finds (1) the selectivity filter, (2)
a hydrophobic cavity and (3) the intracellular gate. The selectivity
filter (see Fig.~\ref{fig:structures-side}a) is formed by a conserved
Thr-Val-Gly-Tyr-Gly motif. The backbone carbonyls of this motif and
the $\mathrm{O_{\gamma}}$-atoms of the Thr side chains form five
cubic cages each of which is able to coordinate one potassium ion.
The structure of the selectivity filter found in KcsA is conserved
even in eukaryotic channels. That is why KcsA acts as a general model
system that is used to study potassium channel function. 

Many channels can open and close their pore \emph{via} a conformational
change. This so-called gating takes place in a controlled way and
can be provoked by the interaction of the pore-forming protein domain
with other domains, other molecules or in response to electric forces.\citep{Alberts:TheCell:08:11,Yellen:Nature:02}
In the KcsA channel and its homologs, gating can take place \emph{via}
the intracellular gate or \emph{via} conformational changes in the
selectivity filter.\citep{Hoshi:JGenPhysiol:13} Here we investigate
the motions of the filter and their influence on conductance. The
intracellular gate remains in the open state.

We reanalyze the non-equilibrium molecular dynamics simulation data
of the KcsA channel protein that were previously published by Köpfer
et al\emph{.} \citep{Koepfer:Science:2014} and consists of a total
amount of $15.1\,\mu\mathrm{s}$ of MD simulation in 20 short trajectories
with individual lengths ranging between $541.4\,\mathrm{ns}$ and
$793.5\,\mathrm{ns}$. In their simulations, a steady potassium ion
current is maintained by the computational electrophysiology approach
of Kutzner et al\emph{.} \citep{Kutzner:BiophysJ:11}. The simulations
are therefore intrinsically non-reversible and the applications of
methods that were developed for reversible dynamics, like TICA, is
not justified.

In the following, we compute the VAMP components, define long-lived
states in this space using PCCA+ and characterize the thus-obtained
states.

\paragraph{Dynamic modes of the selectivity filter and surrounding residues}

We compute the leading singular functions of the Koopman operator
that describe the KcsA dynamics at a lag time of $40\,\mathrm{ns}$
using VAMP. The input features (ansatz functions) $\boldsymbol{\chi}(t)$
for VAMP consist of two groups: a) all inverse pairwise distances
between heavy atoms of the selectivity filter (residues 75 to 79 in
the first subunit using the numbering scheme of PDB file 1K4C \citep{Zhou:Nature:01}
and their corresponding residues in the other three subunits). b)
the inverse distance to the closest potassium cation for every heavy
atom in the selectivity filter. This results in a total number of
$7750$ features. In the computation of distances, atoms that are
symmetric under a rotation of the side chain dihedral by $\pi$ are
treated as one atom. In this analysis this applies to the atom pairs
$(\mathrm{C}_{\delta1},\,\mathrm{C}_{\delta2})$ and $(\mathrm{C}_{\epsilon1},\,\mathrm{C}_{\epsilon2})$
in tyrosine residues, $(\mathrm{C}_{\gamma1},\,\mathrm{C}_{\gamma2})$
in valine residues, the pair $(\mathrm{O}_{\delta1},\mathrm{O}_{\delta2})$
in aspartic acid residues and the pair $(\mathrm{O}_{\epsilon1},\,\mathrm{O}_{\epsilon2})$
in glutamic acid residues. 

We discarded the first $18\,\mathrm{ns}$ of every trajectory. That
is because in the first $18\,\mathrm{ns}$, we observed conformational
changes at the N-terminal end of the intracellular gate. We see these
conformational changes at the beginning of every trajectory. We suspect
that this might be due the pulling procedure that was used to prepare
the open-gate conformation in Ref. \citep{Koepfer:Science:2014}.

\begin{figure}[!h]
\begin{centering}
\includegraphics{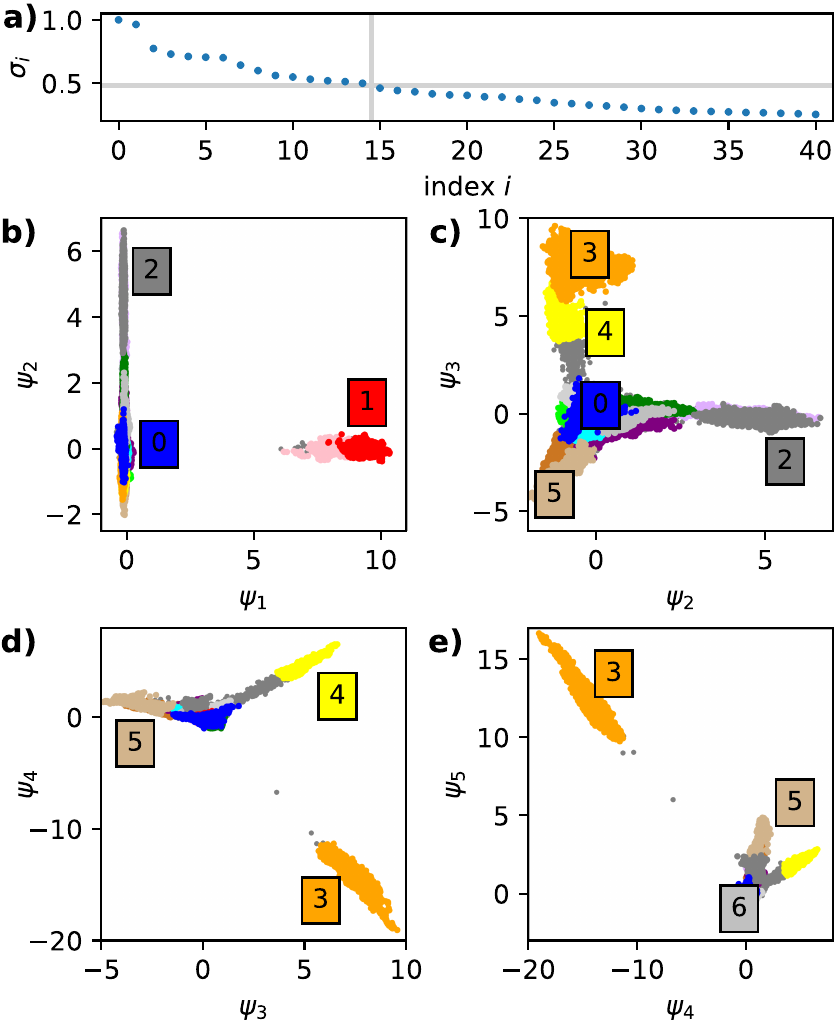}
\par\end{centering}
\caption{\label{fig:KcsA-VAMP}a) Leading 40 singular values obtained with
VAMP. b)-e) Projection of the simulation data on pairs of singular
functions (VAMP components). Data points were colored according to
the macro-state to which they have the highest membership. Data points
that do not clearly belong to any of the macro-states (maximum membership
to any state $<0.6$) are shown as small gray points.}
\end{figure}

The spectrum of singular values (Fig.~\ref{fig:KcsA-VAMP}a) shows
jumps at positions 1, 2, 6, 7, 8 and 14 (not counting the constant
singular value $\sigma_{0}=1$, see Suppl. Fig.~12) and become quasi-continuous
afterwards. We therefore restrict the analysis to the dynamics within
the space of the leading 14 singular functions. To validate this decomposition,
we perform the non-equilibrium Chapman-Kolmogorov test. Results show
(see Suppl. Fig.~13) good agreement between estimates and predictions
for the fast processes (with smaller singular values) and deviations
between predictions and estimates for the slower processes (with large
singular values). As elaborated in the next paragraphs, the KcsA trajectories
contain many unique transition events, which explains the failure
of the dynamic model to provide accurate predictions of the long timescale
kinetics.

\paragraph{Detection of long-lived states with PCCA+}

Projections of the MD data points $\mathbf{x}$ (conformations) onto
pairs of left singular functions $(\psi_{i}(\mathbf{x}),\,\psi_{j}(\mathbf{x}))$
show that the data points form clearly separated clusters (see Fig.~\ref{fig:KcsA-VAMP}b-e).
Such clustering has been observed for many other molecular systems
and indicates the presence of long-lived states.\citep{Weber:ZIB:04}
This motivates us to group the conformation into a small number of
macro-states.

We assign the data points to 15 macro-states using the PCCA+ algorithm.
We apply the PCCA+ variant of Ref.\citep{Weber:ZIB:04} to the data
points in the space of the leading 14 singular functions (see methods
subsection \ref{subsec:clustering} and appendices \ref{subsec:inner-simplex},
\ref{subsec:macrostates}). We observe that the macro-states defined
with PCCA+ match well with the ``density blobs'' that one would
assign intuitively by looking at the projections (see Fig.~\ref{fig:KcsA-VAMP}b-e).
This indicates that the space of the singular functions is a suitable
space for clustering with PCCA+.

\paragraph{Transitions between long-lived states and their populations}

\begin{figure}[!h]
\centering{}\includegraphics{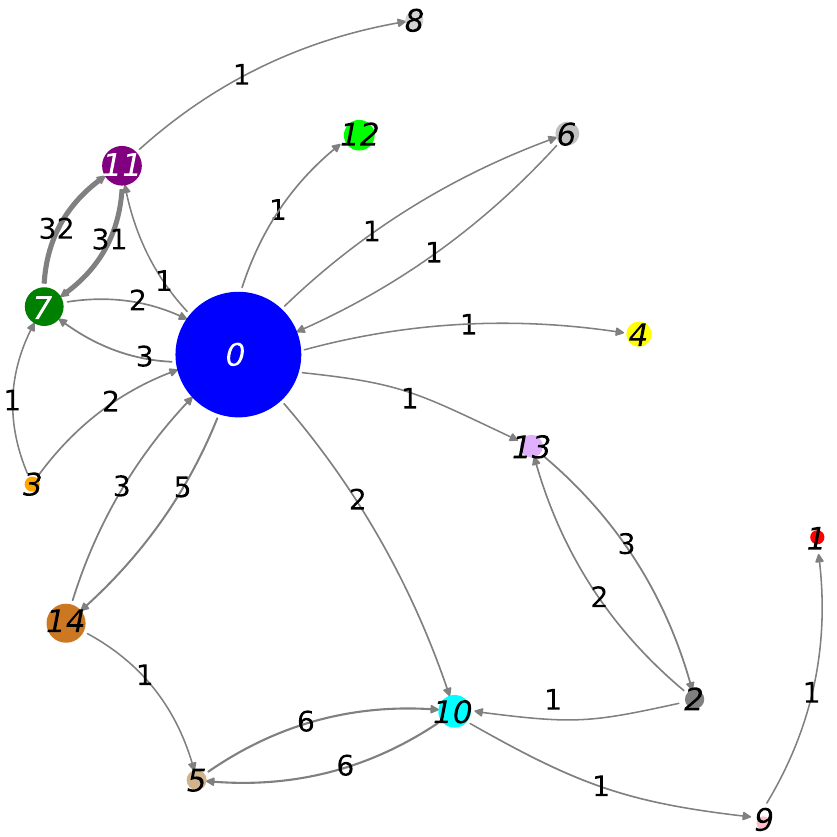}\caption{\label{fig:connectivity-network} Connectivity network for the 15
long-lived states that were identified with VAMP and PCCA+. Long-lived
states are shown as disks with areas proportional to the frequency
of the state in the MD simulation data. Macro-states that are kinetically
connected by transitions in the data are connected by an arrow in
this figure. Numbers on the arrows denote the number of transition
events observed in the MD data. Numbers inside the disks are state
labels.}
\end{figure}

We compute the number of transitions between the macro-states using
the mile-stoning method (also called transition-based assignment or
core set approach, see appendix \ref{subsec:milestoning} and Refs.\citep{Buchete:JPhysChemB:08,Schutte:JChemPhys:11,Jain:JChemTheoryComput:12}).
The network of transitions between the macro-states (Fig.~\ref{fig:connectivity-network})
shows that most transitions occur only once. States 0, 6, 7, 11 and
14 are in the reversibly connected set (ergodically visited macro-states).
Most of the simulation data is assigned to macro-state number 0 (see
Fig.~\ref{fig:connectivity-network}). 

The present MD simulation data does not allow to make any statements
about asymptotic state occupancies in the steady state equilibrium
that might possibly be reached at 100s of microseconds and above.
Most conformational changes observed in the MD data occur only in
one direction. This might indicate a lack of sampling of state transitions
in the short MD data and longer MD simulation might reveal that the
transitions are in fact reversible.

Inspection of the trajectories shows that transition between the cores
can take relatively long (e.g., see the transition from macro-state
0 to macro-state 14 in Fig.~\ref{fig:time-traces-example}). For
some of the states, the transition in/out of the state can take roughly
the same amount of time that the system spends in the state. This
may indicate either that the description of the dynamics requires
more macro-states or that the approximation of the singular functions
with VAMP is not accurate enough. (That is, there exists a better
approximation that would lead to a more metastable kinetics of the
reduced model.)

Ion permeation (except for the blocked states) is faster than the
life-times of the macro-states. The time between ion transition events
is typically on the order of $10\,\mathrm{ns}$ while dwell times
of the macro-states are typically on the order of $100\,\mathrm{ns}$
(see Suppl. table 2 and for example permeation in macro-state 5 and
14 in Fig.~\ref{fig:time-traces-example}). Therefore transitions
between macro-states do not seem to describe the individual ion movement
steps in the permeation mechanism. Rather the macro-states appears
more related to the protein conformation (see next sections).

\paragraph{Long-lived macro-states differ in the occupancy of the selectivity
filter}

\begin{figure}[!h]
\begin{centering}
\includegraphics{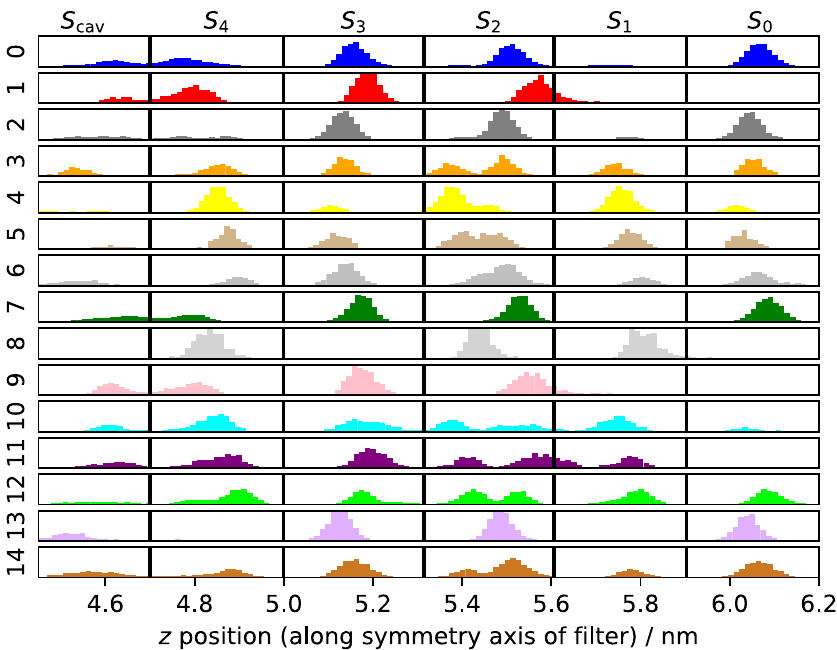}
\par\end{centering}
\caption{\label{fig:ion-histograms}Histogram of ion occupancy along the channel
pore for each macro-state. Black vertical lines mark the positions
of the carbonyl oxygens that demarcate the ion binding sites $\mathrm{S}_{\mathrm{cav}}$,
$\mathrm{S}_{4}$, $\mathrm{S}_{3}$, $\mathrm{S}_{2}$, $\mathrm{S}_{1}$
and $\mathrm{S}_{0}$. In state 8 (light gray), the ion binding site
$\mathrm{S}_{3}$ is occupied by a water molecule. In states 1 (red),
9 (pink) and 10 (cyan) the ion binding site $\mathrm{S}_{1}$ is occupied
by water.}
\end{figure}

We compute histograms of the ion occupancy of the selectivity filter
(Fig.~\ref{fig:ion-histograms}). One histogram is computed for each
macro-state separately. The most frequent state (0, blue), and states
2, 7 and 13 have an evacuated ion binding site $\mathrm{S}_{1}$ that
is neither occupied by a potassium ion nor by a water molecule (see
Suppl. Fig.~6). This means that ions do a long jump from $\mathrm{S}_{2}$
to $\mathrm{S}_{0}$ during conduction in these macro-states and $\mathrm{S}_{1}$
is only visited transiently, with a dwell time that is much shorter
than the dwell time in $\mathrm{S}_{2}$ and $\mathrm{S}_{0}$ (see
for instance the part of the trajectory that is assigned to core 0
in Fig.~\ref{fig:time-traces-example} and Suppl. Figs. 7 to 10).
In other macro-states, e.g. state 11 (violet), $\mathrm{S}_{1}$ is
more frequently occupied. States 1 (red) and 9 (pink) show an ion
binding site $\mathrm{S}_{1}$ that is occupied with water. Furthermore,
these states a characterized by a flipped Tyr78 conformation and a
drastically distorted selectivity filter (see Fig.~\ref{fig:structures-side}.1
and next section). No ion permeation events are observed in these
states (see below). In state 10 (cyan) $\mathrm{S}_{1}$ is partially
occupied by water. Still, this state is conductive. In state 8 (light
gray), the ion binding site $\mathrm{S}_{3}$ is occupied by a water
molecule. No ion conduction takes place in this state (see below).

\paragraph{Long-lived macro-states differ in ion conduction}

\begin{figure}[!h]
\centering{}\includegraphics{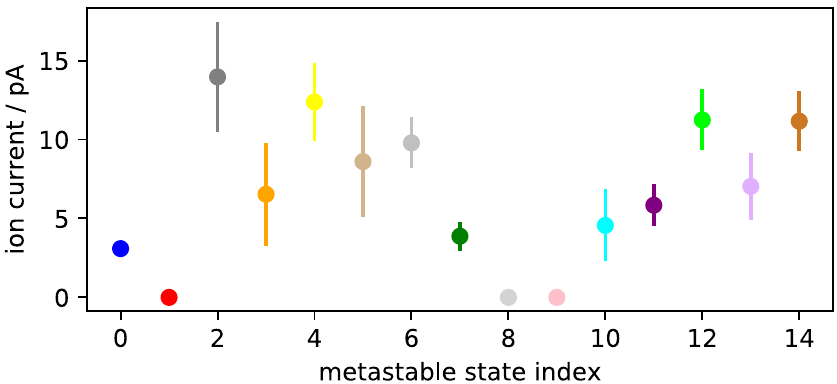}\caption{\label{fig:currents}Potassium ion current for each macro-state. Error
bars show standard deviations and were computed by assuming that the
number of permeation evens is Poisson-distributed (see appendix \ref{subsec:ion-current}).
In states 1, 8 and 9, no permeation events took place. (States 1 and
9 are in the Tyr78-flipped conformation. In state 8, a water molecule
is located in $\mathrm{S}_{3}$.) In the most frequent state 0, the
ion current is relatively low.}
\end{figure}

\begin{figure}[!h]
\begin{centering}
\includegraphics{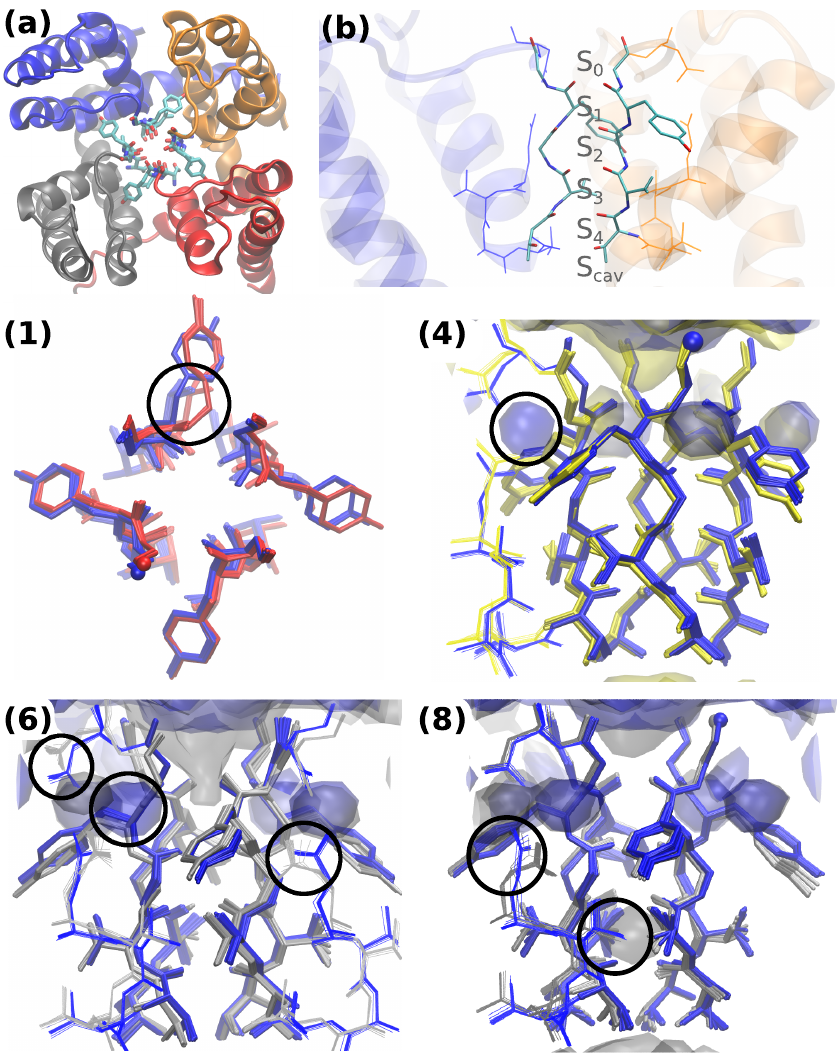}
\par\end{centering}
\caption{\label{fig:structures-side}(a) KcsA seen from the extracellular side,
atoms of the selectivity filter are shown as sticks. (b) Cross-section
through the filter (side view). Filter atoms are shown as sticks,
surrounding atoms are shown with lines. Ion binding sites are labeled
$\mathrm{S}_{\mathrm{cav}}$ through $\mathrm{S}_{0}$. (1) View of
the selectivity filter from the extracellular side. Structures (conformations)
in blue are drawn from macro-state 0. Superimposed on that are structures
from macro-state 1. A remarkable deviation from state 0 in the Tyr78
conformation is marked with a black circle. (4), (6), (8): Side view
of the selectivity filter and surrounding amino acids. Structures
(conformations) in blue are drawn from macro-state 0. Superimposed
on that are structures from different macro-states (4, 6, 8). 20 structures
are shown per macro-state where every conformations is an average
over 50 conformations drawn randomly from the macro-state. Residues
forming the pore of the selectivity filter are shown as sticks, residues
surrounding the filter (including Glu71 and Asp80) with thin lines.
Water density is shown as semi-transparent isosurfaces. The tiny blue
sphere marks the oxygen atom in Gly79 of the first subunit of the
channel. Significant deviations (compared to state 0) in the Glu71-Asp80
contact and buried water presence are indicated with black circles.
All states except 0 show disruption of the Glu71-Asp80 interaction
in one or two subunits. This disruption can be accompanied by the
absence of a buried water molecule like in state 4 (yellow), state
6 (light gray) and state 5 (tan, not shown).}
\end{figure}

\begin{figure*}
\includegraphics[width=1\textwidth]{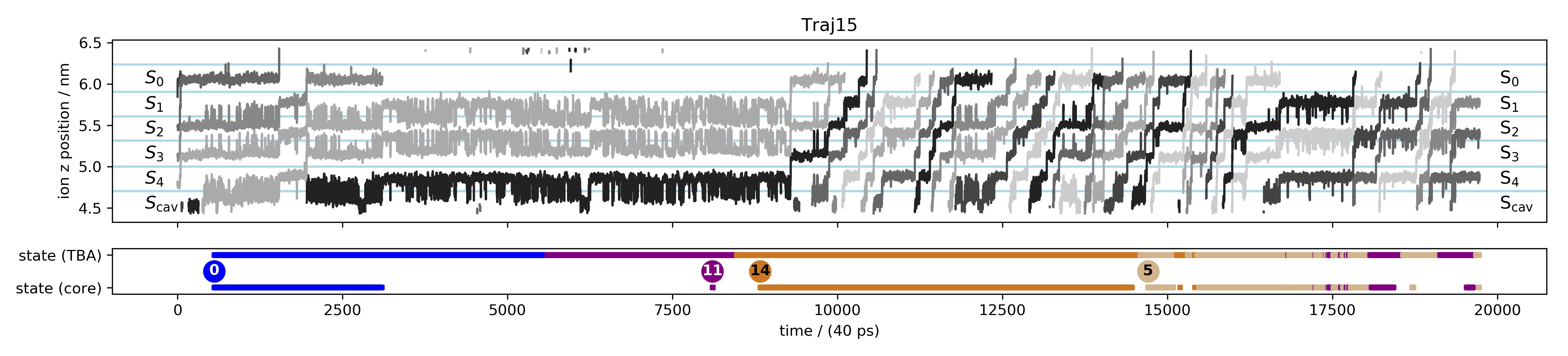}

\caption{\label{fig:time-traces-example}Exemplary time series of potassium
ion positions and assignment to metastable states. A pair of plots
is shown, which share the time axis. The top plot shows the $z$ positions
of all ions in the selectivity filter. The bottom plot shows the metastable
state visited at time $t$. Metastable states are color-coded and
the index of every state is shown in a circle on the first core entry.
Two variants (core-based assignment and transition-based assignment,
TBA) of metastable state assignments are shown. The trajectory of
cores only shows frames where the conformation can be assigned with
a high probability (membership) to a macro-state. Frames that were
left unassigned in the core trajectory are assigned to the most recent
or most proximate core in the TBA trajectories by splitting transitions
at the midpoint.\citep{Buchete:JPhysChemB:08}}
\end{figure*}

We compute the potassium ion current through the pore by counting
the net number of forward ion transitions from the $\mathrm{S}_{1}$
to the $\mathrm{S}_{0}$ binding site (see appendix~\ref{subsec:ion-current}
for details). The ion current for each macro-state is shown in Fig.~\ref{fig:currents}.
Computing the ion current from the transitions from $\mathrm{S}_{4}$
to $\mathrm{S}_{3}$ gave identical results. Ion current differs significantly
between macro-states. However it should be noted that there is only
little simulation data for the different macro-states. For many states
less than 20 permeation events are observed (see also Suppl. table~2).
In states 1 (red), 8 (light gray) and 9 (pink) no permeation is observed.
In states 1 and 9, the filter is in the Tyr78-flipped conformation.
In state 8 a water molecule was threaded into the ion file. The most
frequently visited state in the MD simulation (state number 0) conducts
little compared to most other states. Since the free energies of the
states cannot be computed from the present simulation data, no statement
can be made about the contribution of the different states to the
overall conduction.

\paragraph{Differences in Tyr78 conformation, Glu71-Asp80 interaction and presence
of buried water}

To characterize the structural features of the macro-states, we sampled
randomly with replacement 1000 conformations from every state. Average
conformations for every macro-state are shown in Fig.~\ref{fig:structures-side}.

In states 1 (red) and 9 (pink) Tyr78 is in a flipped conformation
(compared to to state 0). The Tyr78 flip coincides with a disruption
of the selectivity filter structure and zero conduction. 

The Glu71-Asp80 contact \citep{Cordero:NatStructMolBiol:06} can either
be formed, or it can be broken in one or in two subunits of the channel.
It is formed in all subunits in state 0. It is open in one subunit
in states 2, 3, 4, 7, 8, 10, 12, and 14. The contact is open in two
subunits in states 1, 5, 6, 9, 11, and 13 (see Suppl. table 1). Opening
and closing of the Glu71-Asp80 contact is a reversible process (since
there are reversible transitions between states 0, and 6, 7, 11, 14).
We observe opening of two Glu71-Asp80 contacts only in adjacent subunits
of the channel which may hint to cooperativity. 

In contrast to what is known about the KcsA channel,\citep{Cordero:NatStructMolBiol:06}
opening of the Glu71-Asp80 contact in these MD simulation does not
inactivate the channel. Instead ion current in the open-contact states
is slightly increased (exceptions: 1, 8, 9; see explanations above)
over the ion current in the closed-contact state 0. 

The water molecule that is involved in the Glu71-Asp80 interaction
can either be present or absent.  We observe at most one absent water
molecule. Breaking of the Glu71-Asp80 interaction does not strictly
coincide with the absence of water.

In state 2 we observe a highly tilted Glu71 side chain conformation
where the side chain points toward the filter pore. The tilt is larger
than in the crystal structures and NMR structures of the KcsA protein
that are available in the Protein Data Bank. This conformation might
be stabilized by electrostatic attraction between the carboxyl groups
of Glu71 and the potassium ions. 

In summary our analysis of the MD data of Köpfer et al. \citep{Koepfer:Science:2014}
reveals 15 long-lived states. While the most frequent state shows
a crystal-like conformation of the selectivity filter other states
show flipping of the Tyr86 side chain or opening of the Glu71-Asp80
contact in one or more subunits. The different identified states display
distinct ion conductances, establishing a direct link between channel
function and the the conformations identified with VAMP and PCCA+.

\section{Conclusion }

We have used the Variational Approach to Markov Processes (VAMP) to
formulate a dynamical dimension reduction method for identifying the
collective variables of the ``slow'' or ``rare'' processes in
many-body systems. In this formulation, VAMP can replace the TICA
method that is only defined for statistically reversible and stationary
dynamics, and in practice often only usable when the probability distribution
sampled by the simulation trajectories is close to equilibrium.

We have applied VAMP-based dimension reduction to the asymmetric simple
exclusion process toy model for single file ion diffusion and to non-equilibrium
molecular dynamics data of the KcsA potassium channel protein. Both
systems have high-dimensional state spaces and follow non-equilibrium
dynamics that do not comply with the principle of detailed balance
(microscopic reversibility). For both systems, we could construct
a low-dimensional model that captures physically interesting processes.

We have demonstrated that VAMP is superior to TICA in correctly estimating
kinetic distances for the intrinsically non-reversible ASEP model.
Based on theoretical insights, we expect this to be true for any non-reversible
system. For the analysis of non-equilibrium data that originates from
simulating a reversible system with a non-equilibrium initial condition,
we empirically showed that TICA and VAMP give similar results.

We have shown that the space of the leading singular functions is
a suitable space for identification of long-lived macro-states even
for the case of non-reversible dynamics. This was confirmed twice
for the KcsA protein data: (1) the PCCA+ macro-states appear as well-separated
density-clusters when projected to the space of the singular functions,
(2) counting exit end entry events with the core-set (or transition-based)
approach confirms that transitions between macro-states are rare events.

We proposed to extend the scope of the Chapman-Kolmogorov test from
an application to probabilities \citep{Prinz:JChemPhys:11} to general
observables. We further proposed to use the singular functions as
observables for the Chapman-Kolmogorov test. In fact it has been shown
that the singular functions span the space of indicator functions
for \emph{coherent sets} \citep{Koltai:Computation:18}. Coherent
sets are particular stable sets in time-space.\citep{Banisch:Chaos:17}
Examples for coherent sets are oceanic \citep{Froyland:PhysRevLett:07,Banisch:Chaos:17}
or atmospheric \citep{Santitissadeekorn:PhysRevE:10} eddies. Reliably
simulating their formation and dissolution should be equally challenging
as sampling the exit from metastable states in systems with reversible
dynamics. Testing whether a reduced dynamical model captures these
rare events seem worthwhile.

\paragraph{Software availability}

The linear Variational Approach to Markov Processes has been implemented
in the publicly available PyEMMA software package \href{http://emma-project.org}{http://emma-project.org}.

\paragraph{Acknowledgements}

F.P. acknowledges funding from the Yen Post-Doctoral Fellowship in
Interdisciplinary Research and from the National Cancer Institute
of the National Institutes of Health (NIH) through Grant CAO93577.
F.P. and F.N. acknowledge funding by European Commission (ERC CoG
772230 ``ScaleCell''), Deutsche Forschungsgemeinschaft (SFB1114/C03).
F.N. acknowledges funding by MATH+ (AA1-6). H.W. acknowledges funding
from 1000-Talent Program of Young Scientists in China. B.d.G. and
F.N. gratefully acknowledge financial support from the Deutsche Forschungsgemeinschaft
(FOR 2518 ``DynIon''). We are grateful to Brooke E. Husic and Erik
H. Thiede for their corrections to the manuscript, to Simon Olsson
for the intensive exchange about the modeling of many-particle systems
and to Cecilia Clementi for insightful discussions about the ASEP
model.

\appendix

\section{Proof of \eqref{eq:kinetic-map}\label{sec:Proof-of-kinetic-map}}

The SVD of $\mathcal{K}_{\tau}$ is
\begin{equation}
\mathcal{K}_{\tau}g=\sum_{i}\sigma_{i}\langle g,\phi_{i}\rangle_{\rho_{1}}\psi_{i},
\end{equation}
then the transition density can be expressed as
\begin{eqnarray}
p(\mathbf{z}|\mathbf{x}) & = & \mathcal{K}_{\tau}\delta_{\mathbf{z}}(\mathbf{x})\\
 & = & \sum_{i}\sigma_{i}\psi_{i}(\mathbf{x})\phi_{i}(\mathbf{z})\rho_{1}(\mathbf{z}).
\end{eqnarray}
Considering the orthonormality of singular functions, we have

\begin{multline}
D_{\tau}^{2}(\mathbf{x},\mathbf{y})=\int\frac{\left(\sum_{i}\sigma_{i}\left(\psi_{i}(\mathbf{x})-\psi_{i}(\mathbf{y})\right)\phi_{i}(\mathbf{z})\rho_{1}(\mathbf{z})\right)^{2}}{\rho_{1}(\mathbf{z})}\,\mathrm{d}\mathbf{z}\\
=\sum_{i,j}\sigma_{i}\sigma_{j}\left(\psi_{i}(\mathbf{x})-\psi_{i}(\mathbf{y})\right)\left(\psi_{j}(\mathbf{x})-\psi_{j}(\mathbf{y})\right)\left\langle \phi_{i},\phi_{j}\right\rangle _{\rho_{1}}\\
=\sum_{i}\sigma_{i}^{2}\left(\psi_{i}(\mathbf{x})-\psi_{i}(\mathbf{y})\right)^{2}.
\end{multline}

\section{Models}

\subsection{Koopman matrix for the ASEP model\label{sec:ASEP-model} }

Let $\wedge$ denote the bitwise \noun{and} operator. Let $\mathbf{L}\in\mathbb{R}^{2^{N}\times2^{N}}$.
For all $0\leq i<2^{N}$, $0\leq j<2^{N}$, $i\neq j$, let
\begin{align}
L_{ij} & =\alpha\;\text{ if }i\wedge1=0\text{ and }j\wedge1=1\\
L_{ij} & =\beta\;\text{ if }i\wedge2^{N-1}=1\text{ and }j\wedge2^{N-1}=0\\
L_{ij} & =p\;\text{ if }\exists\,0\leq k<N-1:i\wedge2^{k}=1\text{ and }\nonumber \\
 & i\wedge2^{k+1}=0\text{ and }j\wedge2^{k}=0\text{ and }j\wedge2^{k+1}=1\\
L_{ij} & =q\;\text{ if }\exists\,0\leq k<N-1:i\wedge2^{k}=0\text{ and }\nonumber \\
 & i\wedge2^{k+1}=1\text{ and }j\wedge2^{k}=1\text{ and }j\wedge2^{k+1}=0\\
L_{ij} & =0\;\text{ otherwise}
\end{align}
and $L_{ii}=-\sum_{j\neq i}L_{ij}$.

The model transition matrix $\mathbf{T}_{\tau}\in\mathbb{R}^{2^{N}\times2^{N}}$
is computed by taking the matrix exponential of $\tau\mathbf{L}$
where $\tau$ is the lag time. The full Koopman operator $\mathcal{K}_{\tau}$
is finite-dimensional for this model and is identical to $\mathbf{T}_{\tau}$

\section{Methods}

\subsection{Implementation of the non-equilibrium Chapman-Kolmogorov test\label{subsec:CK-test-implementation}}

In the Chapman-Kolmogorov test, the estimate of the time-lagged cross-correlation
$\mathrm{cov}_{\mathrm{est}}(f,g;\,n\tau)$ and its model prediction
$\mathrm{cov}_{\mathrm{pred}}(f,g;\,n\tau)$ are compared.

Using Eq. \ref{eq:K-SVDd}, it is possible to express the covariance
at the unit lag time $1\times\tau$ as

\begin{align}
\mathrm{cov}_{\mathrm{pred}}(f,g;\,\tau)= & \mathrm{cov}_{\mathrm{est}}(f,g;\,\tau)\\
= & \langle f,\mathcal{K}_{\tau}g\rangle_{\rho_{0}}\\
= & \sum_{i}\langle g,\phi_{i}\rangle_{\rho_{1}}\sigma_{i}\langle\psi_{i},f\rangle_{\rho_{0}}\\
= & \mathbf{q}^{\top}\mathbf{r}
\end{align}
where we have defined $q_{i}:=\langle g,\phi_{i}\rangle_{\rho_{1}}$
and $r_{i}:=\sigma_{i}\langle\psi_{i},f\rangle_{\rho_{0}}$. 

By combining Eqs. \ref{eq:K-SVDd} and \ref{eq:Koopman-MarkovProperty},
the prediction for higher multiples $n>1$ of the lag time of can
be computed as

\begin{equation}
\mathrm{cov}_{\mathrm{pred}}(f,\,g;n\tau):=\langle f,\mathcal{K}_{\tau}^{n}g\rangle_{\rho_{0}}=\mathbf{q}^{\top}\mathbf{P}^{n-1}\mathbf{r}\label{eq:VAMP-cov}
\end{equation}
where $P_{ij}:=\sigma_{i}\langle\psi_{i},\phi_{j}\rangle_{\rho_{1}}$.
The quantities $\mathbf{P}$, $\mathbf{q}$, and $\mathbf{r}$ can
all be computed from the data and from the spectral quantities that
VAMP provides an approximation for. 

\subsection{Computing the metastable memberships\label{subsec:inner-simplex}}

We use the ``inner simplex'' algorithm of the PCCA+ method \citep{Weber:ZIB:02}
to compute the linear map $\mathbf{A}$ from the space of singular
functions to the space of macro-state memberships. The ``inner simplex''
algorithm was motivated by the observation that reversible metastable
systems show a clustering of data points close to the $N$ most distant
points in the space of the dominant $N$ eigenfunctions (counting
the constant eigenfunction). This so called ``simplex structure''
forms the basis for many spectral clustering algorithms.\citep{Weber:ZIB:02} 

The algorithm in the version of Ref. \citep{Weber:ZIB:02} consist
of two stages: 1) localizing the $N$ most distant points (the \emph{vertices})
$\{\boldsymbol{\psi}_{1}^{\mathrm{(ex)}},\ldots,\boldsymbol{\psi}_{N-1}^{\mathrm{(ex)}}\}$
in the $N-1$-dimensional space of the dominant eigenfunctions (excluding
the constant eigenfunction) and 2) computing barycentric coordinates
for every point $\boldsymbol{\psi}(t)$ with respect to the vertices
by solving the following equations for $m_{i}(t)$
\begin{align}
\boldsymbol{\psi}(t) & =\sum_{i}^{N}m_{i}(t)\boldsymbol{\psi}_{i}^{\mathrm{(ex)}}\\
1 & =\sum_{i}^{N}m_{i}(t).
\end{align}
The solution of this linear problem implicitly defines the linear
map $\mathbf{A}$ from $\boldsymbol{\psi}$ to $\mathbf{m}$.

If the data points $\{\boldsymbol{\psi}(t)\}_{t}$ indeed form clusters
close to the vertices, the coefficient $m_{i}(t)$ can be understood
as the membership of point $\boldsymbol{\psi}(t)$ in the macro-state
number $i$. Here we apply the ``inner simplex'' algorithm not in
the space of the eigenfunctions of the MSM transition matrix as initially
suggested in Ref.\citep{Weber:ZIB:02} but in the space of the singular
functions.

Note that in its conventional use, PCCA+ is applied to cluster the
space of MSM eigenvectors. For MSMs, only one representative value
of the eigenfunctions is needed for every micro-state, because the
approximations to the eigenfunctions are constant on every micro-state
by definition. This is different in our application of PCCA+ to the
continuous order parameters computed with VAMP. Hence, all data points
have to be used here.

\subsection{Defining macro-states and core sets\label{subsec:macrostates}}

In order to interpret the macro-states that originate from VAMP and
PCCA+, we investigate representative molecular conformations from
every macro-state. To this, we first define the core set $C_{i}$
of the macro-state number $i$ as

\begin{equation}
C_{i}:=\{\mathbf{x}(t)\mid i=\mathrm{argmax}_{j}m_{j}(t)\text{ and }m_{i}(t)\geq f\}
\end{equation}
where $0.5\leq f\leq1$ is some arbitrary cut-off on the memberships.
We chose $f=0.6$. From each core set, we draw 1000 random samples
of molecular conformations with replacement. A subset of these conformations
are shown Fig.~\ref{fig:structures-side}.

\subsection{Counting transitions and finding the largest connected set of macro-states\label{subsec:milestoning}}

We count transitions at a lag time of $\tau=40\,\mathrm{ps}$ according
to the transition-based assignment (TBA) algorithm or mile-stoning
algorithm.\citep{Schutte:JChemPhys:11,Buchete:JPhysChemB:08} In the
TBA algorithm, every conformation $\mathbf{x}(t)$ is first assigned
to the either the last core set that was hit by the trajectory or
the next core set that will hit by the trajectory, whichever is closer
in time. We thus obtain a sequence of core labels $\{s(t)\}_{t=0,\ldots,T}$,
$s(t)\in\mathbb{N}$ for every MD trajectory. For every trajectory,
we compute a count matrix $\mathbf{c}$ from $\{s(t)\}_{t}$ using
the standard approach \citep{Prinz:JChemPhys:11} as follows
\begin{equation}
c_{ij}=\sum_{t=0}^{T-\tau}\delta_{is(t)}\delta_{js(t+\tau)}
\end{equation}
where $\delta_{ij}$ is the Kronecker delta. The count matrix for
all trajectories $\mathbf{C}$ is computed by summing the individual
count matrices of each trajectory.

The largest connected set of macro-states \citep{Prinz:JChemPhys:11}
is computed from $\mathbf{C}$ and consists of the five states 0,
6, 7, 11 and 14.

\subsection{Computing the ion current\label{subsec:ion-current}}

We estimate the potassium ion current by computing the number of times
some ion transitions from binding site $\mathrm{S}_{1}$ to binding
site $\mathrm{S}_{0}$ of the selectivity filter. We estimate the
number of transitions using a core-set approach.\citep{Buchete:JPhysChemB:08}
The core region of the $\mathrm{S}_{1}$ site is defined as $5.7\,\mathrm{nm}\leq z_{\mathrm{ion}}\leq5.85\,\mathrm{nm}$
and the core region of the $\mathrm{S}_{0}$ size as $6.0\,\mathrm{nm}\leq z_{\mathrm{ion}}\leq6.14\,\mathrm{nm}$
(in the coordinate system of the MD data from Ref.\citealp{Koepfer:Science:2014}).
For all trajectory segments that are assigned to macro-state $i$,
we compute $n_{i}$ the number of transitions from the core of $\mathrm{S}_{1}$
to the core of $\mathrm{S}_{0}$ minus the number of reverse transitions
(summing up the number of transitions of all the ions). The error
$\Delta n_{i}$ of $n_{i}$ is computed as $\sqrt{n_{i}}$ by assuming
that $n_{i}$ is Poisson distributed. The ion current is computed
as $I_{i}=\frac{en_{i}}{\Delta t_{i}}$ where $e$ is the elementary
charge and $\Delta t_{i}$ is the length of the trajectory pieces
assigned to macro-state $i$. The error is estimated as $\Delta I_{i}=\frac{e\Delta n_{i}}{\Delta t_{i}}$.

\bibliographystyle{apsrev4-1}
\bibliography{bib/fab,bib/own,bib/all}

\end{document}

\title{\emph{Supplementary Information}\\
Identification of kinetic order parameters for non-equilibrium dynamics}

\author{Fabian Paul$^{1,2}$, Hao Wu$^{3,1}$, Maximilian Vossel${{}^4}$,
Bert L. de Groot${{}^4}$, Frank Noé$^{1,5*}$}

\maketitle
1: FU Berlin, Department of Mathematics and Computer Science, Arnimallee
6, 14195 Berlin, Germany

2: University of Chicago, 929 East 57th Street Chicago, IL 60637,
USA

3: Tonguing University, School of Mathematical Sciences, Shanghai,
200092, P.R. China

4: Max Planck Institute for Biophysical Chemistry, Am Fassberg 11
D-37077 Göttingen, Germany

5: Rice University, Department of Chemistry, Houston, Texas 77005,
USA

\renewcommand{\figurename}{Suppl. Fig.}

\section{Supplementary figures for the ASEP toy model}

\begin{figure}[H]
\begin{centering}
\includegraphics[width=1\textwidth]{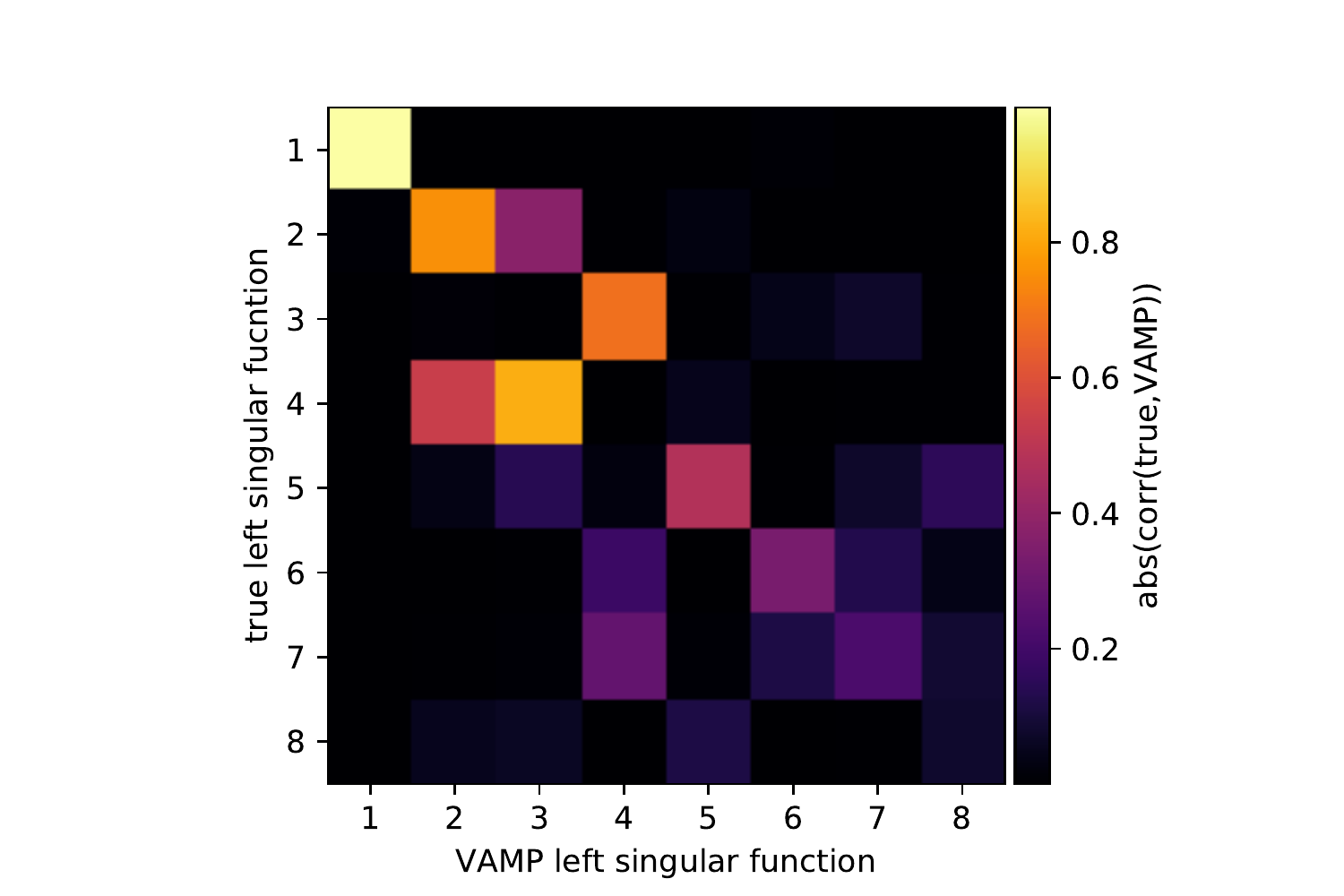}
\par\end{centering}
\caption{Pearson correlations between the true left singular functions computed
from the full transition matrix and the VAMP estimate of the left
singular functions using the basis of occupancy vectors and one-hot
encoded occupancy counts (see main text for details). The limitation
of the basis does not allow to express all the true singular functions
and leads to inconsistent orderings, if the singular functions are
only sorted by the magnitude of the singular values. Counting of singular
functions starts at zero here, so 1 is the first non-constant function.}
\end{figure}
\begin{figure}[H]
\begin{centering}
\includegraphics[width=0.5\textwidth]{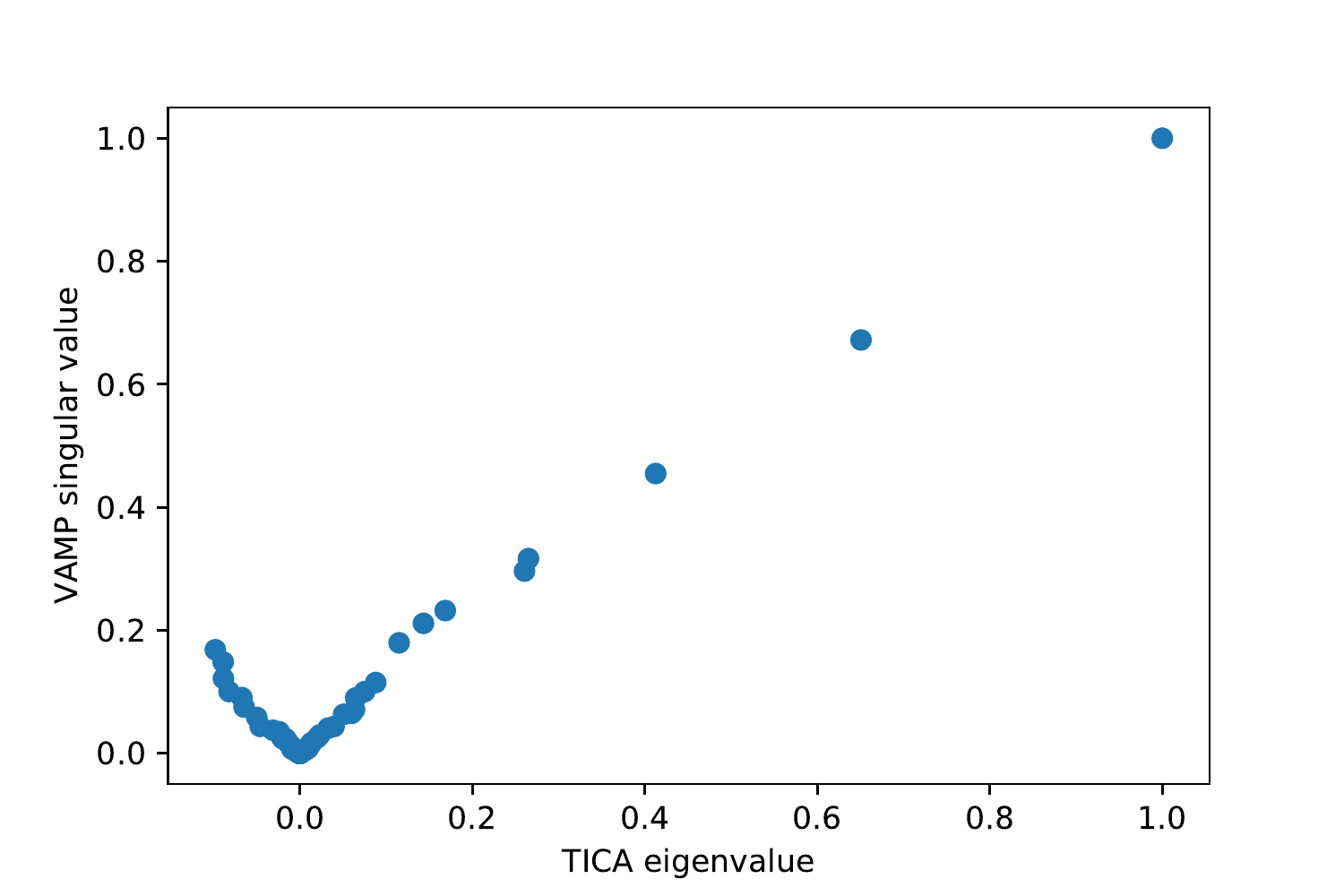}
\par\end{centering}
\caption{Comparison of the singular values computed using VAMP with the eigenvalues
computed with TICA for the ASEP model and using a complete basis.
Spectral components were computed from the true transition matrix
for VAMP and TICA. Results in this figure are therefore unaffected
by sampling error or errors due to an inexpressive basis set.}
\end{figure}

\begin{figure}[H]
\begin{centering}
\includegraphics[width=1\textwidth]{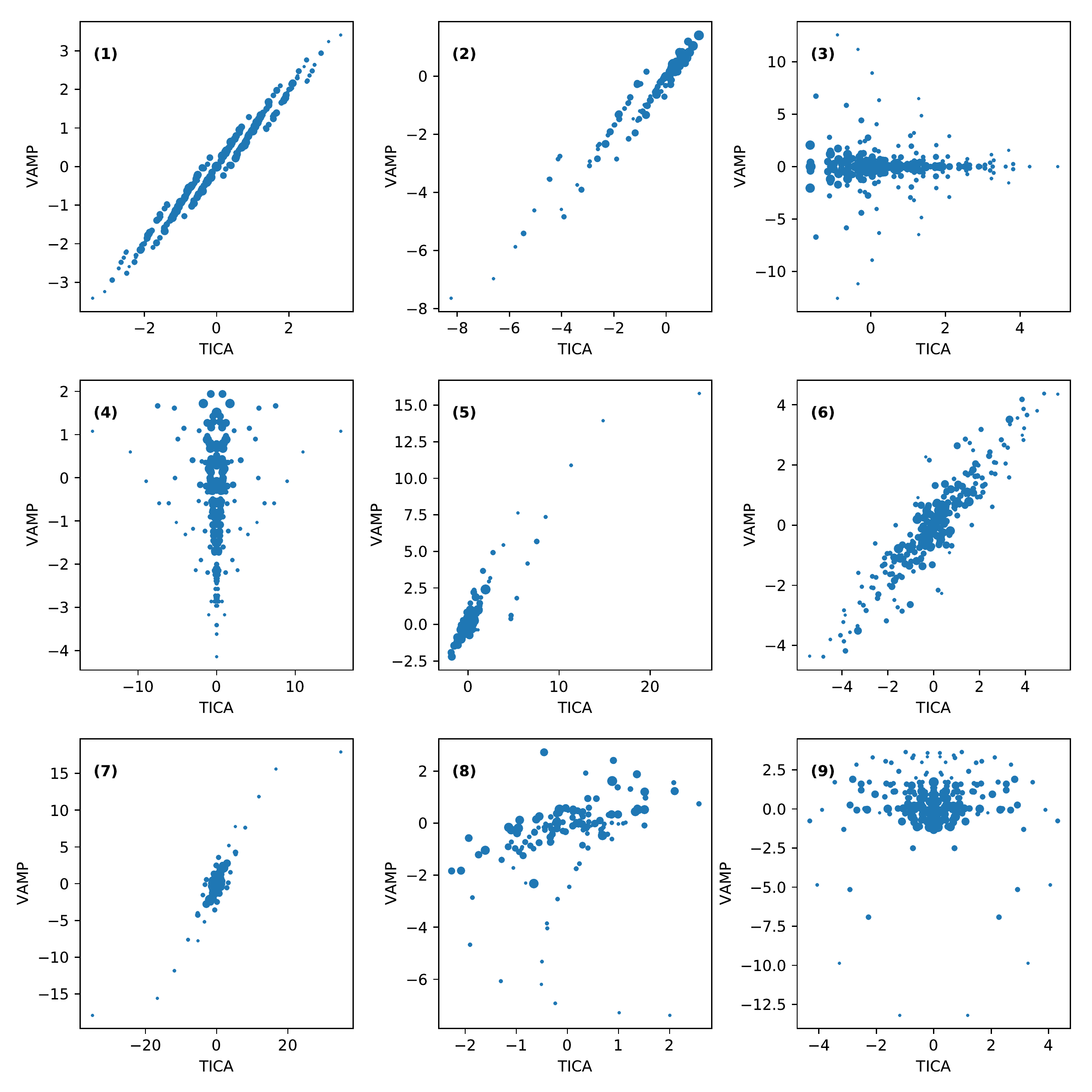}
\par\end{centering}
\caption{Comparison of the singular vectors computed with VAMP with the eigenvectors
computed with TICA for the ASEP model and using a complete basis.
Spectral components were computed from true transition matrix for
VAMP and TICA. Results in this figure are therefore unaffected by
sampling error or errors due to an inexpressive basis set.}
\end{figure}

\begin{figure}[H]
\begin{centering}
\includegraphics[width=0.45\textwidth]{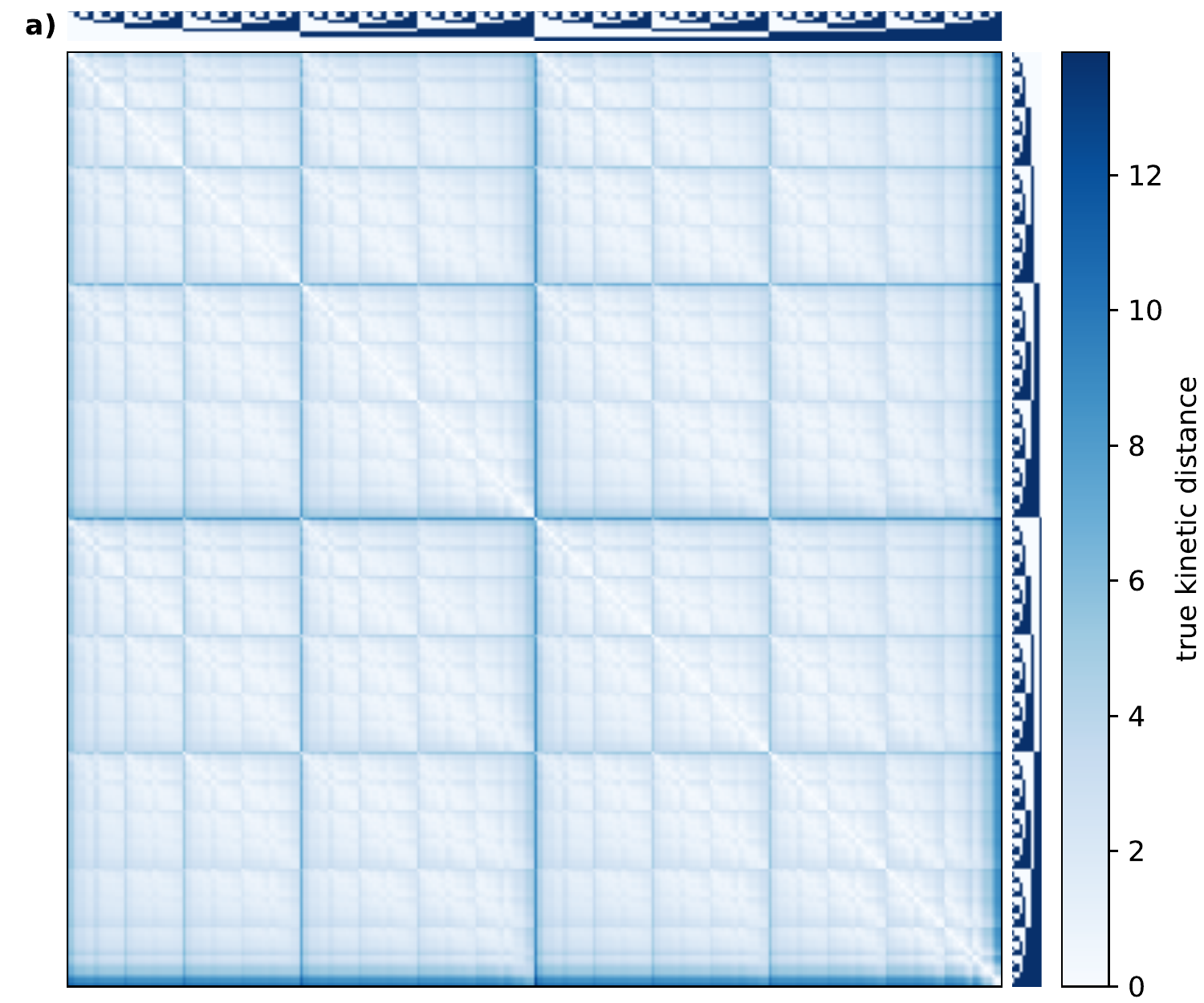}\includegraphics[width=0.45\textwidth]{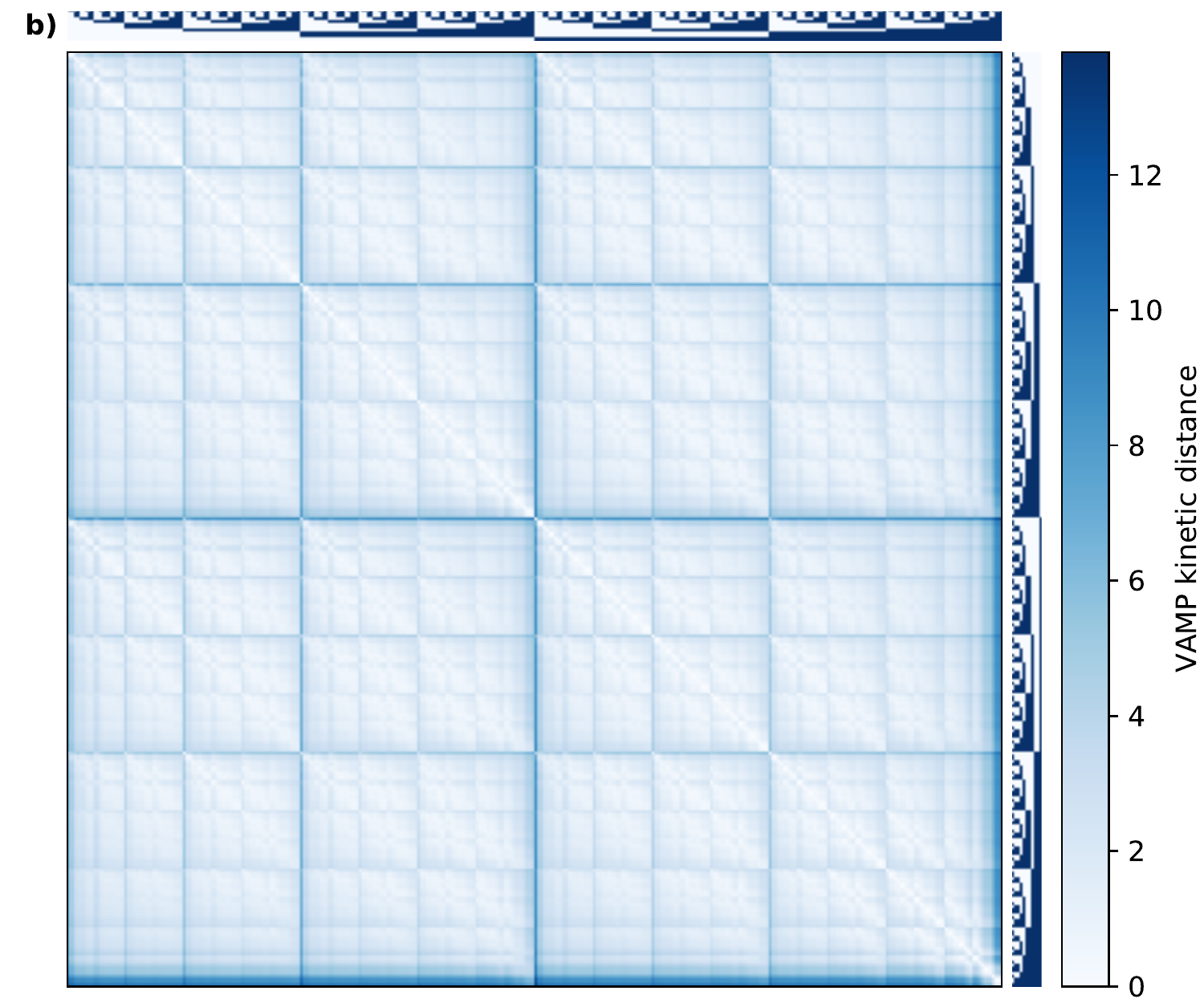}
\par\end{centering}
\begin{centering}
\includegraphics[width=0.45\textwidth]{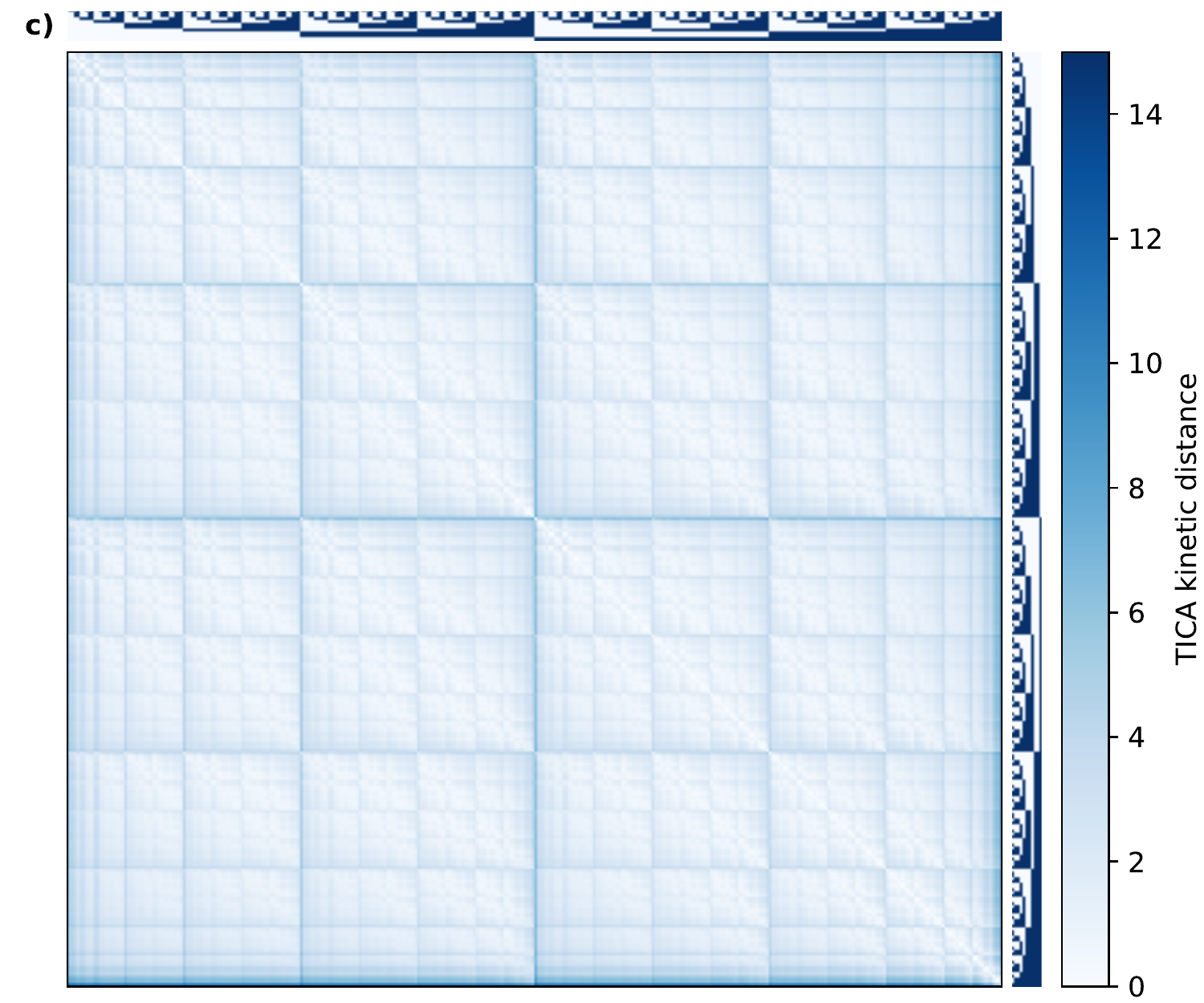}\includegraphics[width=0.45\textwidth]{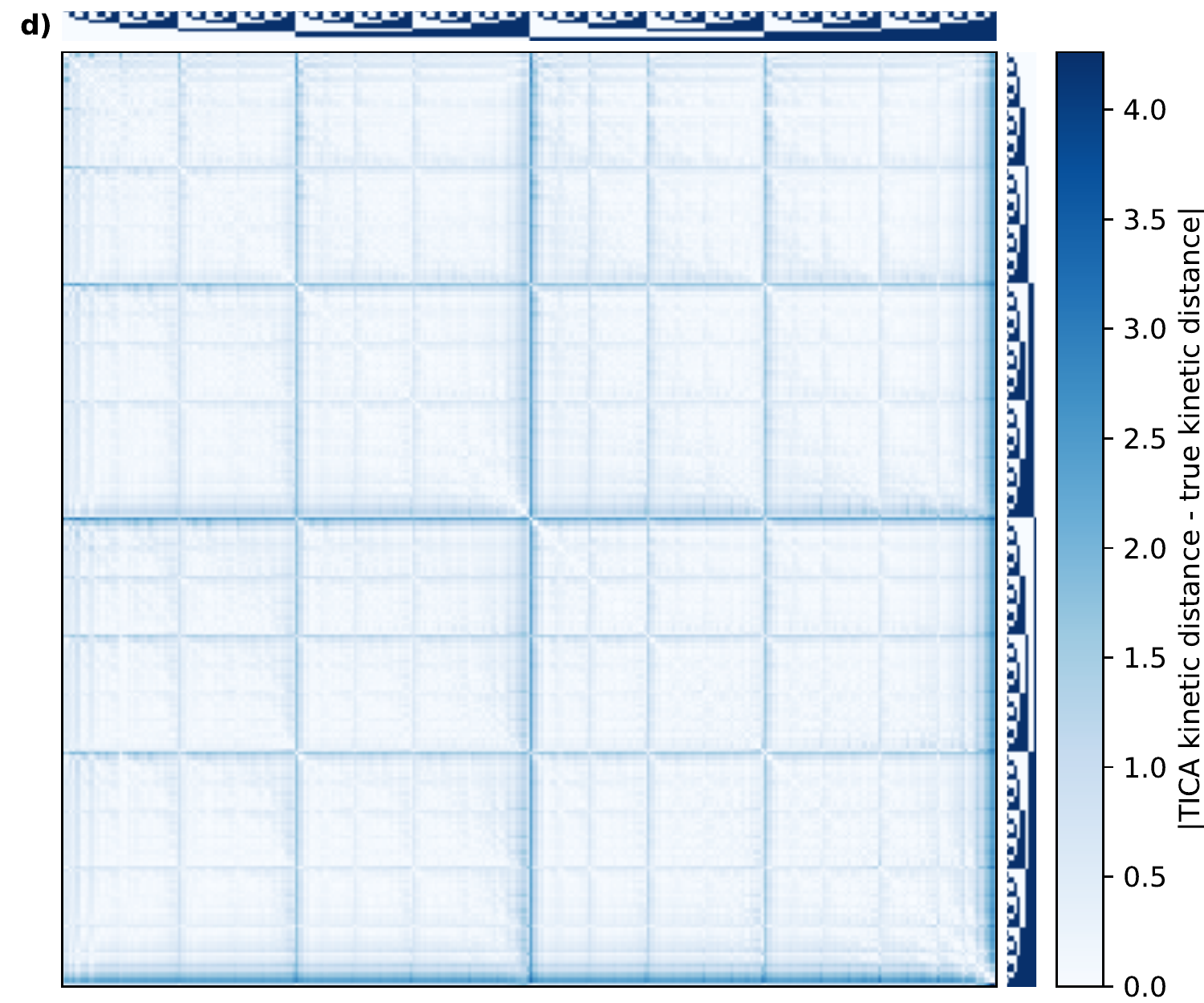}
\par\end{centering}
\begin{centering}
\includegraphics[width=0.45\textwidth]{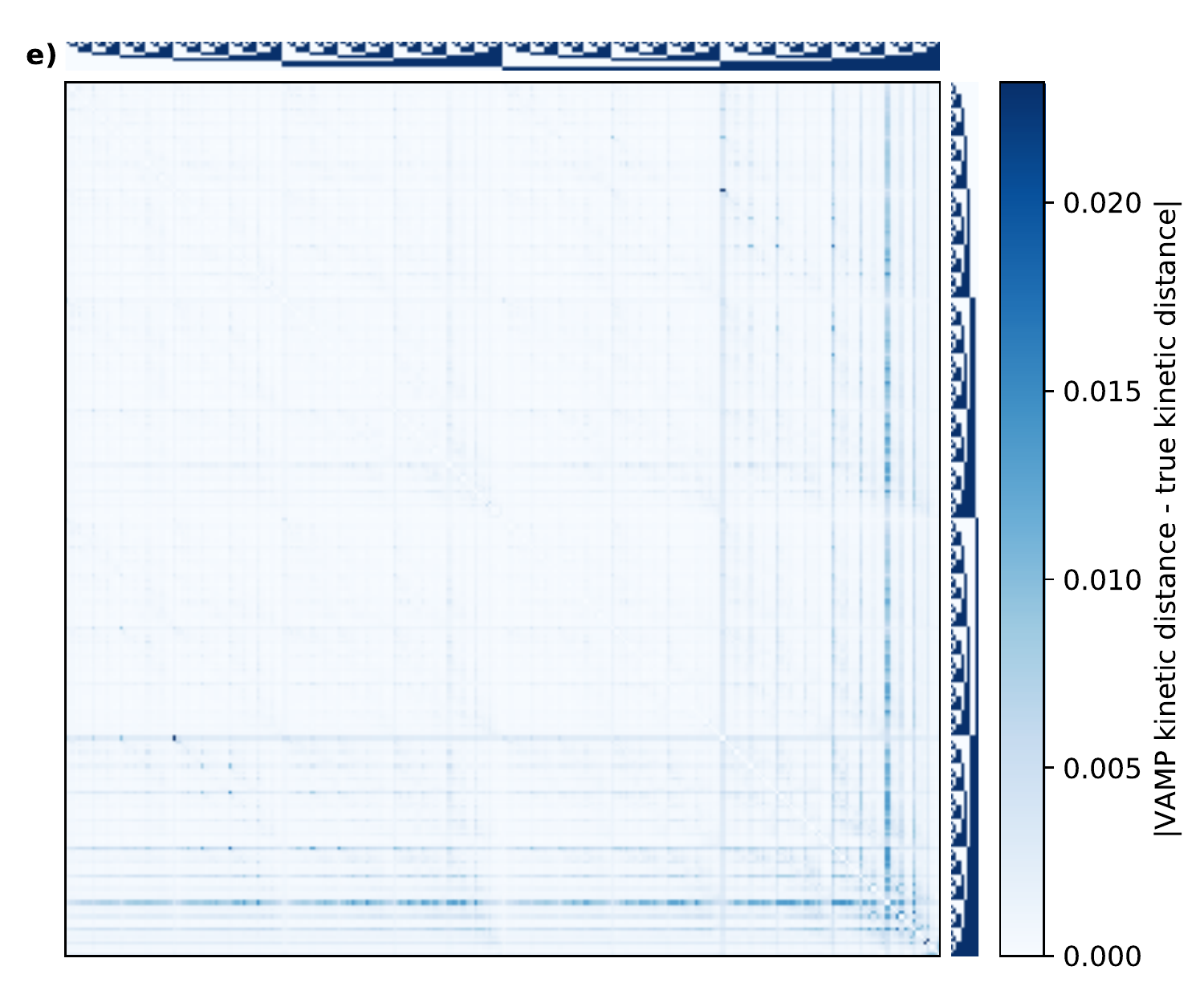}
\par\end{centering}
\caption{\label{fig:kinetic-distance}Comparison of kinetic distance matrices
computed with VAMP and TICA respectively to the true kinetic distances
(true distances were computed directly from the transition matrix
using the defining equation (11) from the main text). The dominant
30 spectral components have been used in the case of VAMP and TICA.
A complete basis and the true transition matrix was used for all computations,
the only difference being the method.}
\end{figure}

\begin{figure}
\begin{centering}
\includegraphics[height=2.5cm]{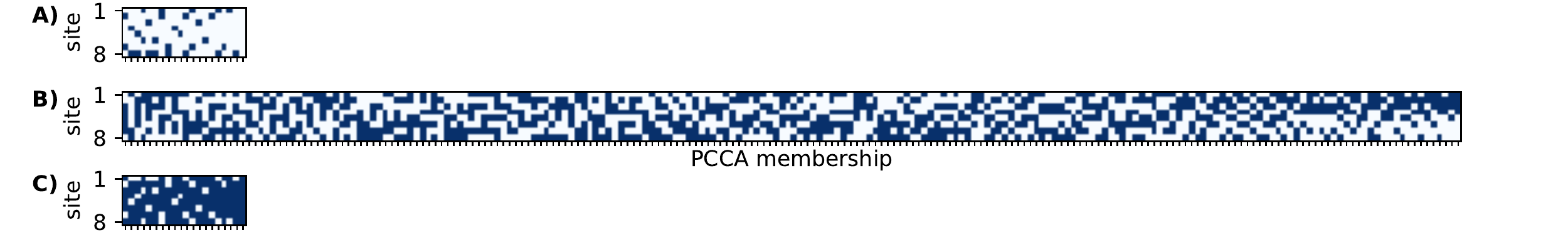}\vspace{1cm}
\par\end{centering}
\begin{centering}
\includegraphics[height=2.5cm]{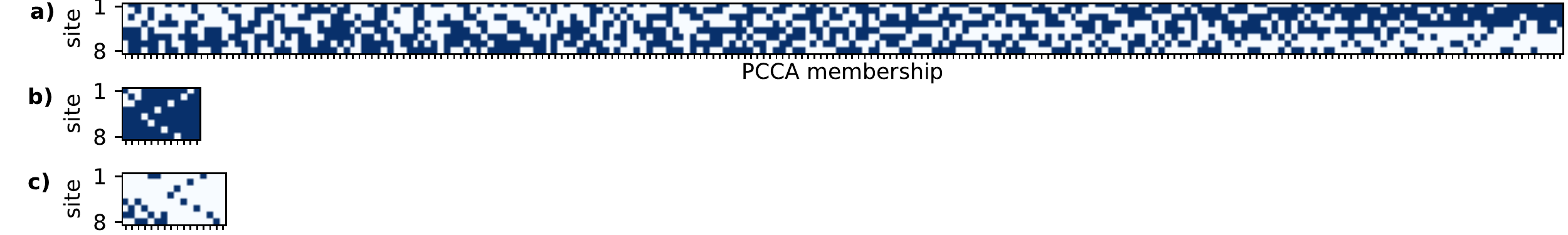}
\par\end{centering}
\caption{Comparison of the three dominant macro-states computed from the true
transition matrix of the ASEP model (A, B, C) and the three dominant
macro-states computed from the low-dimensional model that was estimated
form the simulation data and that uses an incomplete basis of input
features (a, b, c). Every column in each subplot corresponds to a
system state (micro-state) that is represented here using its occupancy
pattern. Dark squares mark occupied sites and light squares mark unoccupied
sites. Micro-states are ordered by increasing macro-state memberships
form the left to the right. We observe qualitative agreement between
the true macro-states and the macro-states approximated from data,
in particular for the high-membership micro-states (right-most columns).
We see an empty state, a full state and a state with a shock (jump
from occupied to unoccupied) in the middle of the queue. If the same
analysis is carried out with more than three states, the results from
the approximated model start to diverge from the true results.}
\end{figure}

\section{Supplementary figures for the KcsA channel protein}

\begin{figure}[H]
\begin{centering}
\includegraphics[width=1\columnwidth]{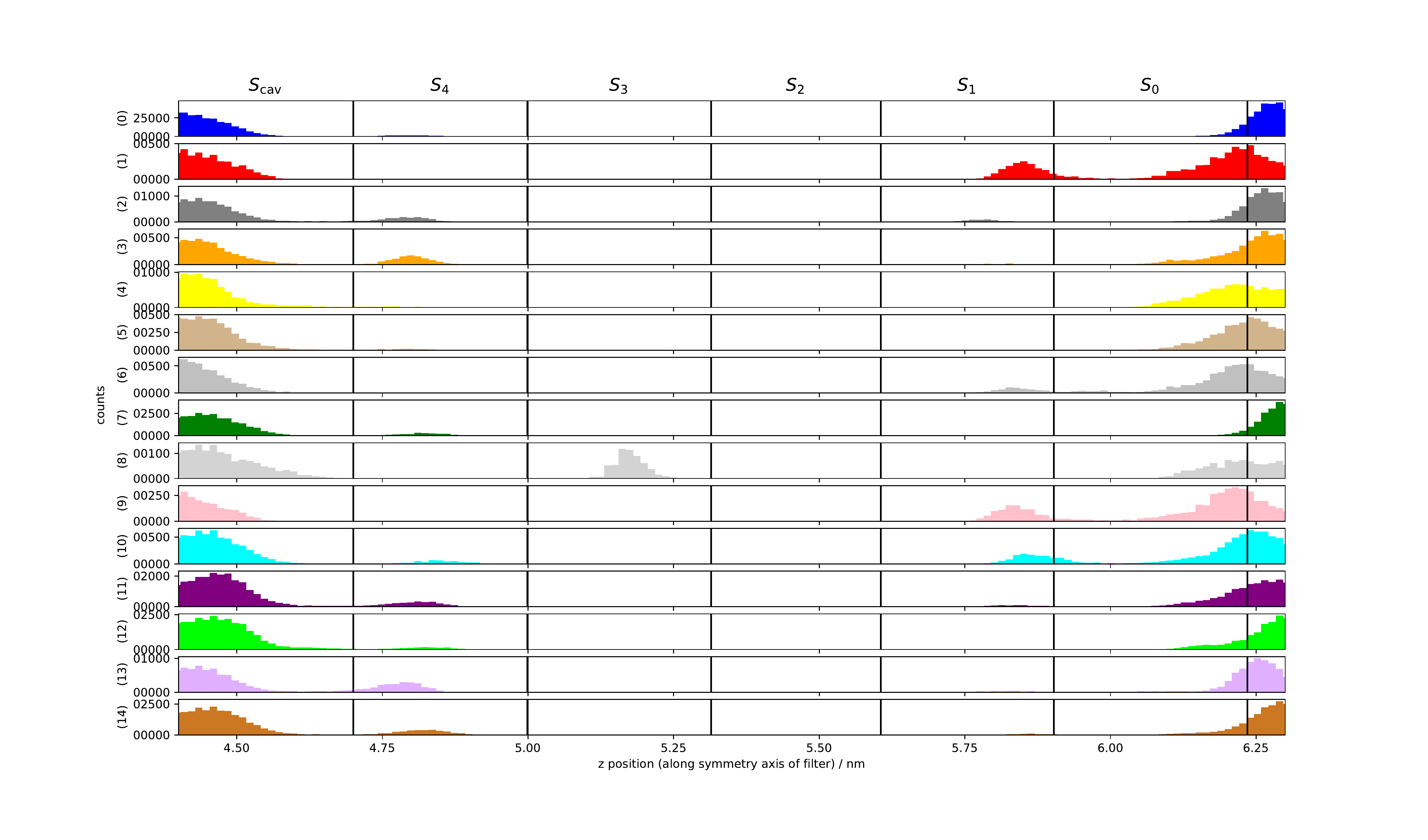}
\par\end{centering}
\caption{\label{fig:water-histograms}Histogram of \emph{water} occupancy along
the channel pore for each metastable state.Black vertical lines mark
the positions of the carbonyl oxygens that demarcate the ion binding
sites $\mathrm{S}_{\mathrm{cav}}$, $\mathrm{S}_{4}$, $\mathrm{S}_{3}$,
$\mathrm{S}_{2}$, $\mathrm{S}_{1}$ and $\mathrm{S}_{0}$. In state
8 (light gray), the ion binding site S3 is occupied by a water molecule.
In states 1 (red), 9 (pink) and 10 (cyan) the ion binding site S1
is occupied by water. }
\end{figure}

\begin{table}[H]
\begin{centering}
\begin{tabular}{|c|c|c|c|c|c|c|c|c|c|}
\hline 
state & current/$\mathrm{\ensuremath{pA}}$ & $\mathrm{S}_{0}$ & $\mathrm{S}_{1}$ & $\mathrm{S}_{2}$ & $\mathrm{S}_{3}$ & $\mathrm{S}_{4}$ & \#open E71-D80 & missing $\mathrm{H}_{2}\mathrm{O}$ & Y78 flipped\tabularnewline
\hline 
\hline 
\color{c0}0 & 3.1 & 0.23 & 0.03 & 0.24 & 0.24 & 0.14 & 0 & 0 & no\tabularnewline
\hline 
\color{c1}1 & 0.0  & 0.01 & 0.07 & 0.23 & 0.30  & 0.23 & 2 & 0 & yes\tabularnewline
\hline 
\color{c2}2 & 13.8 & 0.25 & 0.04 & 0.28 & 0.25 & 0.08 & 1 & 0 & no\tabularnewline
\hline 
\color{c3}3 & 6.7 & 0.15 & 0.12 & 0.27 & 0.17 & 0.14 & 1 & 0 & no\tabularnewline
\hline 
\color{c4}4 & 12.5 & 0.06 & 0.25 & 0.30  & 0.07 & 0.25 & 1 & 1 & no\tabularnewline
\hline 
\color{c5}5 & 9.4 & 0.12 & 0.19 & 0.30  & 0.13 & 0.20  & 2 & 1 & no\tabularnewline
\hline 
\color{c6}6 & 4.6 & 0.05 & 0.18 & 0.23 & 0.19 & 0.20  & 2 & 1 & no\tabularnewline
\hline 
\color{c7}7 & 3.2 & 0.24 & 0.02 & 0.25 & 0.24 & 0.13 & 1 & 0 & no\tabularnewline
\hline 
8 & 0.0 & 0.02 & 0.30  & 0.32 & 0.00 & 0.32 & 1 & 0 & no\tabularnewline
\hline 
\color{c9}9 & 0.0 & 0.01 & 0.05 & 0.25 & 0.30  & 0.17 & 2 & 0 & yes\tabularnewline
\hline 
\color{c10}10 & 6.0 & 0.00  & 0.17 & 0.25 & 0.22 & 0.20  & 1 & 0 & no\tabularnewline
\hline 
\color{c11}11 & 10.6 & 0.11 & 0.17 & 0.29 & 0.14 & 0.18 & 2 & 0 & no\tabularnewline
\hline 
\color{c12}12 & 11.3 & 0.15 & 0.16 & 0.25 & 0.14 & 0.21 & 1 & 0 & no\tabularnewline
\hline 
\color{c13}13 & 7.7 & 0.24 & 0.01 & 0.29 & 0.29 & 0.02 & 2 & 0 & no\tabularnewline
\hline 
\color{c14}14 &  9.5 & 0.19 & 0.08 & 0.26 & 0.21 & 0.12 & 1 & 0 & no\tabularnewline
\hline 
\end{tabular}
\par\end{centering}
\caption{\label{tab:summary}Characterization of metastable states in terms
of selected electrical and structural features. Columns $\mathrm{S}_{0}$
to $\mathrm{S}_{4}$ contain the occupancy probability of the ion
binding sites given that the channel is in one of the metastable states.
The column ``\#open E71-D80'' contains the number of open Glu71-Asp80
contacts (where open is defined as a conformation with maximal oxygen
distance $>0.5\,\mathrm{nm}$). The columns ``missing $\mathrm{H}_{2}\mathrm{O}$''
contains the number of missing buried water molecules near the extracellular
interface of the channel. }
\end{table}

\begin{table}[H]
\begin{centering}
\begin{tabular}{|c|c|c|c|c|}
\hline 
state & \# ion transition events & $\frac{\text{time spent in state}}{\mathrm{ns}}$ & $\frac{\text{time spent in state}}{\mathrm{ns}\cdot\text{ion transtion event}}$  & $\frac{\text{dwell time in state}}{\mathrm{ns}}$ \tabularnewline
\hline 
\hline 
\color{c0}0 & 165 & 8562.0 & 51.9 & 245.1\tabularnewline
\hline 
\color{c1}1 & 0 & 92.7 & NA & 92.7\tabularnewline
\hline 
\color{c2}2 & 16 & 183.6 & 11.5 & 92.9\tabularnewline
\hline 
\color{c3}3 & 4 & 98.1 & 24.5 & 25.5\tabularnewline
\hline 
\color{c4}4 & 25 & 323.6 & 12.9 & 323.6\tabularnewline
\hline 
\color{c5}5 & 6 & 111.8 & 18.6 & 64.5\tabularnewline
\hline 
\color{c6}6 & 37 & 605.8 & 16.4  & 211.5\tabularnewline
\hline 
\color{c7}7 & 18 & 744.3 & 41.3 & 243.4\tabularnewline
\hline 
8 & 0 & 53.4 & NA & 53.4\tabularnewline
\hline 
\color{c9}9 & 0 & 85.7 & NA & 85.7\tabularnewline
\hline 
\color{c10}10 & 4 & 140.5 & 35.1 & 140.5\tabularnewline
\hline 
\color{c11}11 & 19 & 521.6 & 27.4 & 62.5\tabularnewline
\hline 
\color{c12}12 & 34 & 484.6 & 14.2 & 484.6\tabularnewline
\hline 
\color{c13}13 & 11 & 250.7 & 22.8 & 126.3\tabularnewline
\hline 
\color{c14}14 & 34 & 488.2 & 14.3 & 130.6\tabularnewline
\hline 
\end{tabular}
\par\end{centering}
\caption{For all macro-states: summary of ion transition events, time spent
in states, and median dwell time per state. Values in columns 2 and
3 are used to compute the ion current per macro-state.}

\end{table}

\begin{figure}[H]
\includegraphics[width=1\textwidth]{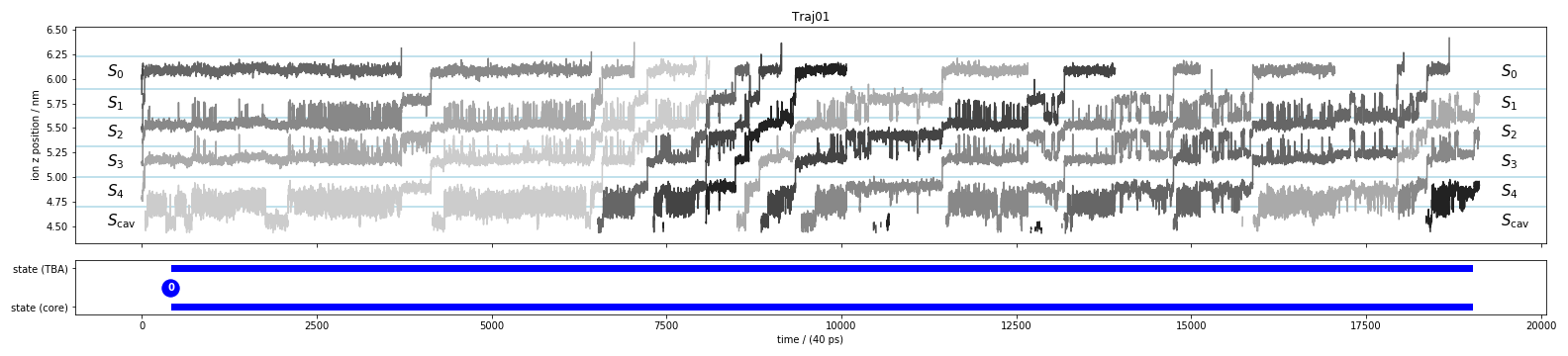}

\includegraphics[width=1\textwidth]{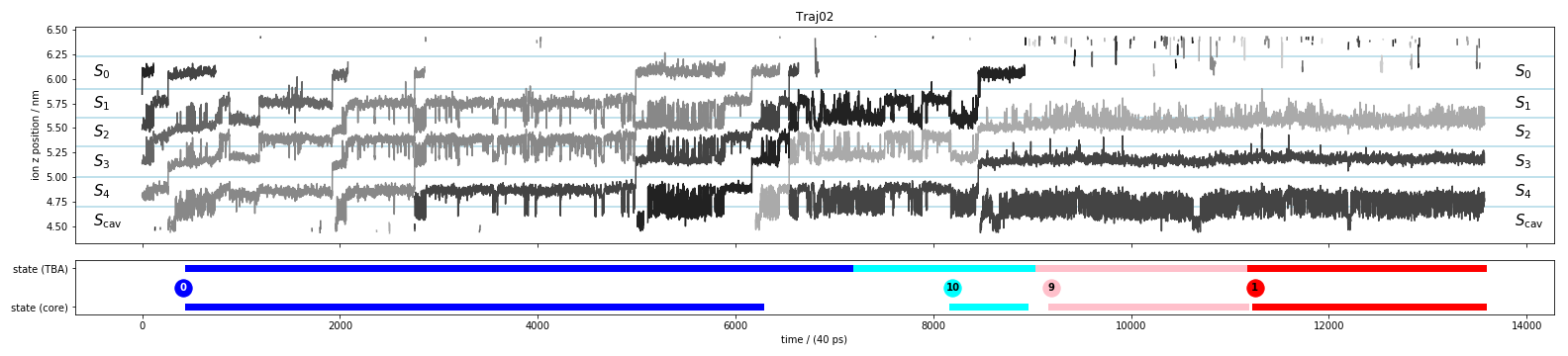}

\includegraphics[width=1\textwidth]{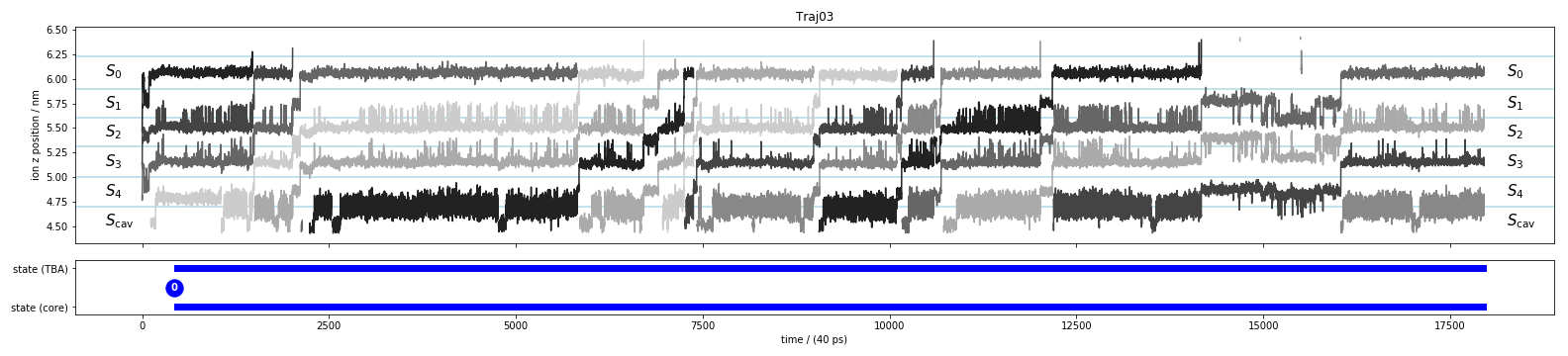}

\includegraphics[width=1\textwidth]{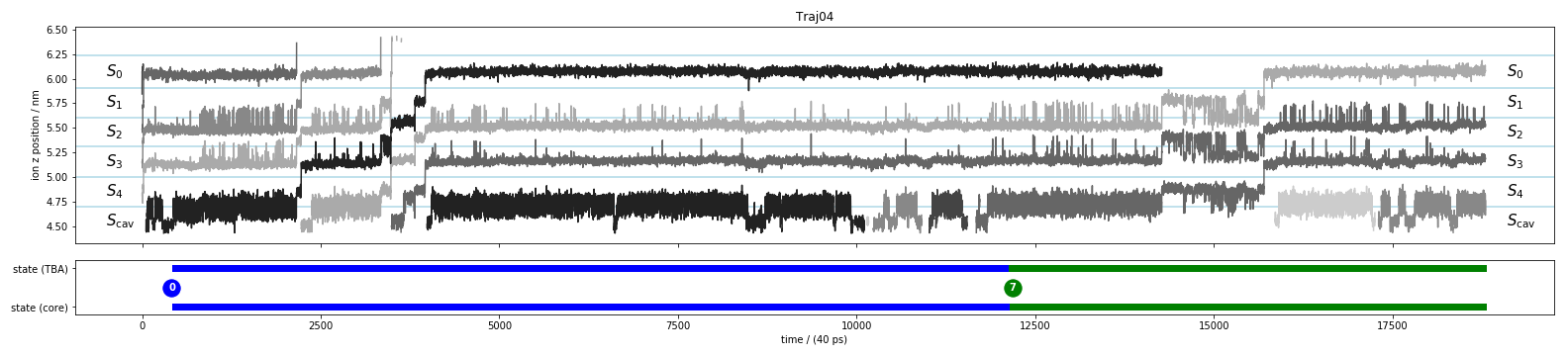}

\includegraphics[width=1\textwidth]{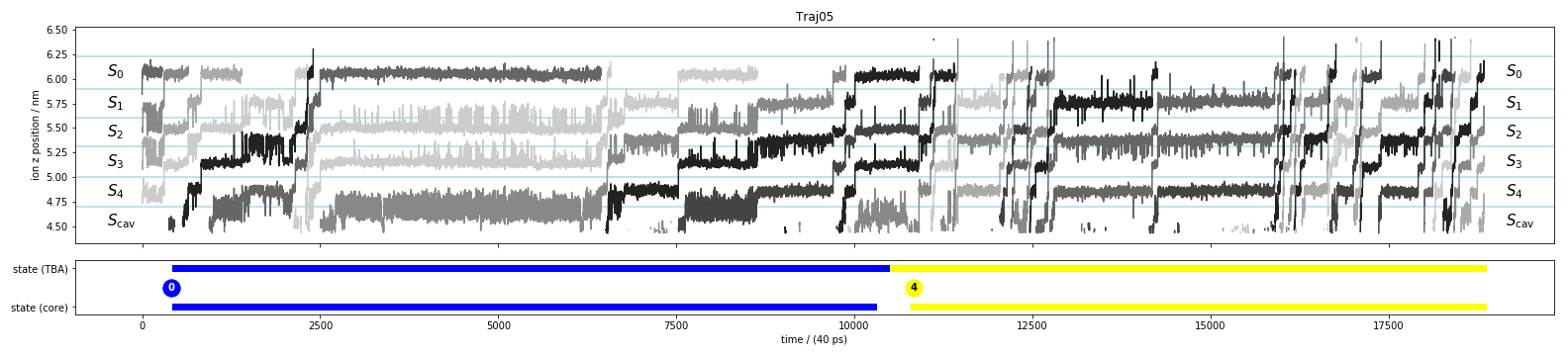}

\caption{\label{fig:time-traces-1}Ion positions and assignment to metastable
state for MD trajectories 1 to 5. For every trajectory, a pair of
plots is shown, which share the time axis. The top plot in every row
shows the $z$ positions of all ions in the selectivity filter. The
bottom plot in every row shows the metastable state visited at time
$t$. Metastable states are color-coded and the index of every state
is shown in a circle on the first state entry in a given trajectory.
Two variants (core-based assignment and transition-based assignment)
of metastable state assignments are shown. The trajectory of cores
only shows frames where the conformation can be assigned with a high
probability (membership) to a metastable state. Frames that were left
unassigned in the core trajectory are assigned to the most recent
or most proximate core in the TBA trajectories by splitting transitions
at the midpoint.}
\end{figure}

\begin{figure}[H]
\includegraphics[width=1\textwidth]{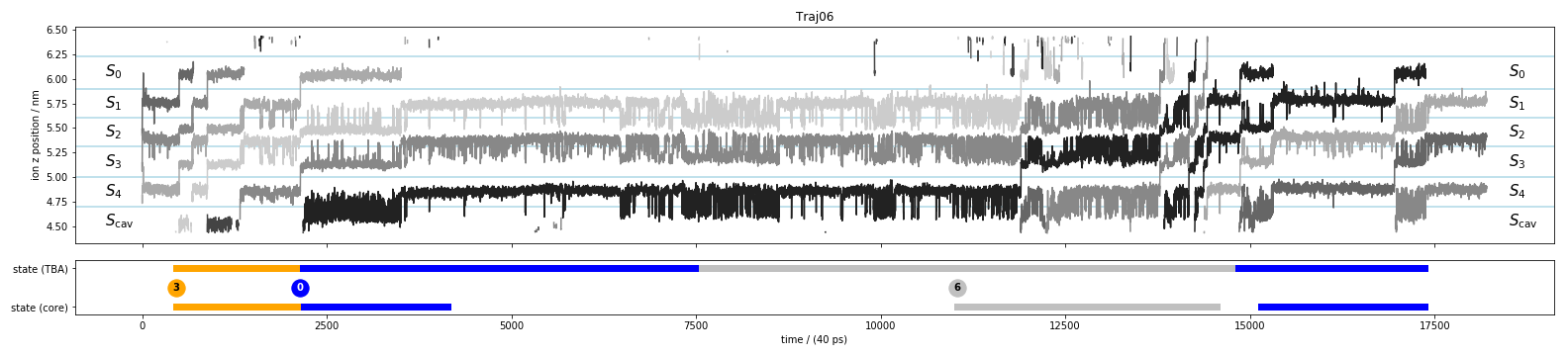}

\includegraphics[width=1\textwidth]{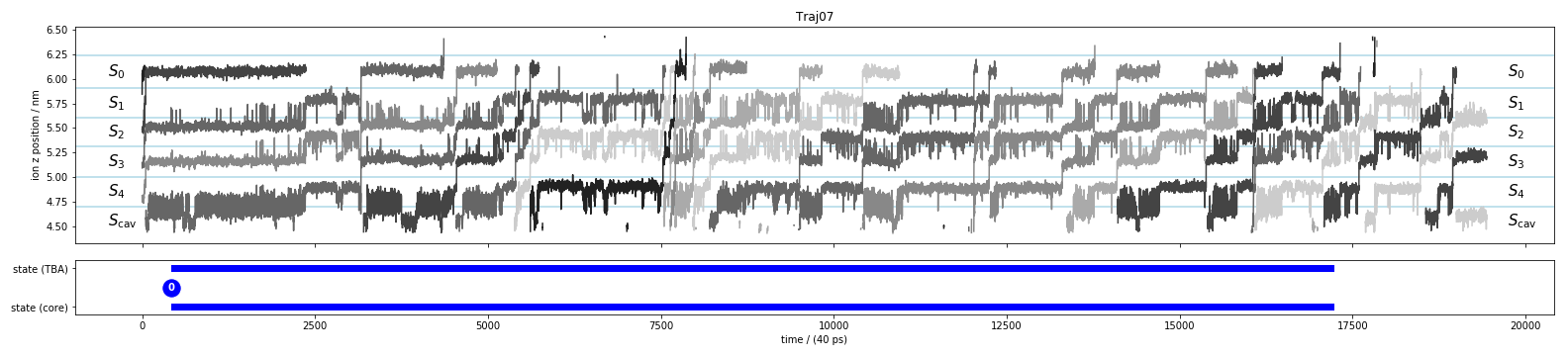}

\includegraphics[width=1\textwidth]{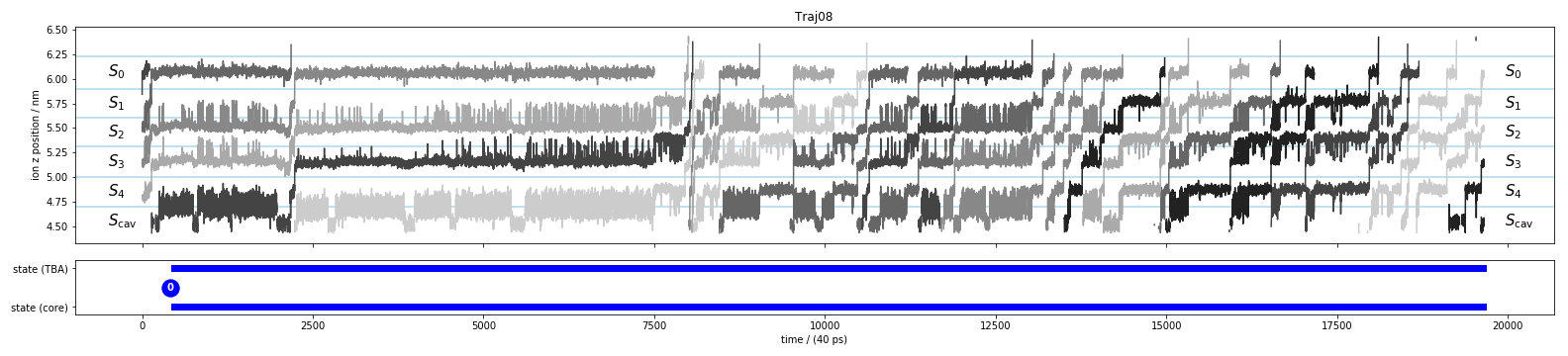}

\includegraphics[width=1\textwidth]{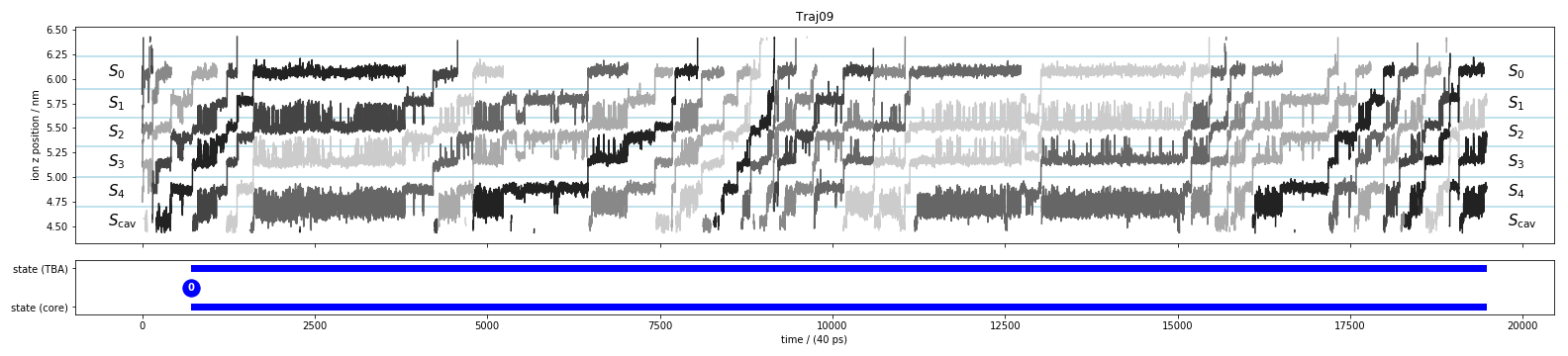}

\includegraphics[width=1\textwidth]{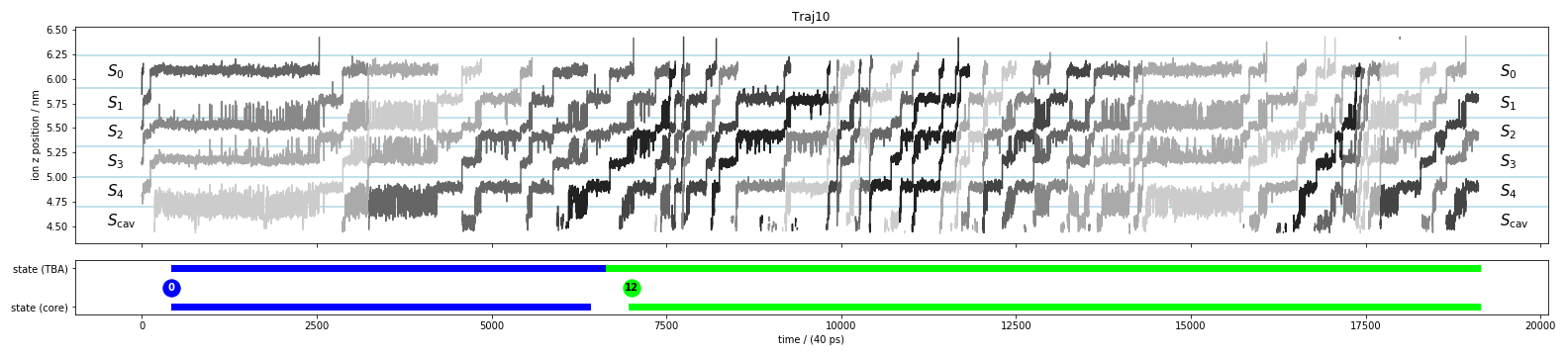}

\caption{\label{fig:time-traces-2}Ion positions and assignment to metastable
state for MD trajectories 6 to 10. For details see caption of figure
\ref{fig:time-traces-1}.}
\end{figure}

\begin{figure}[H]
\includegraphics[width=1\textwidth]{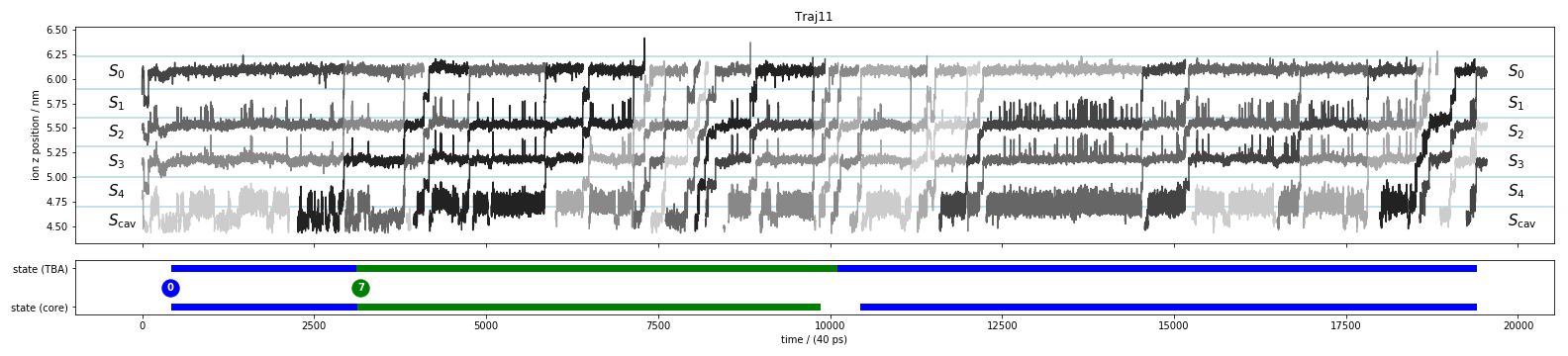}

\includegraphics[width=1\textwidth]{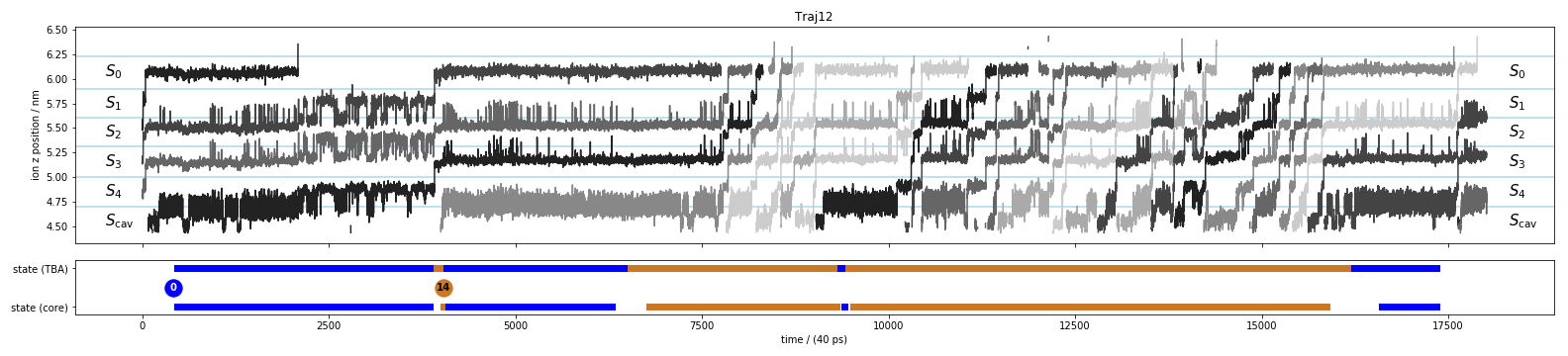}

\includegraphics[width=1\textwidth]{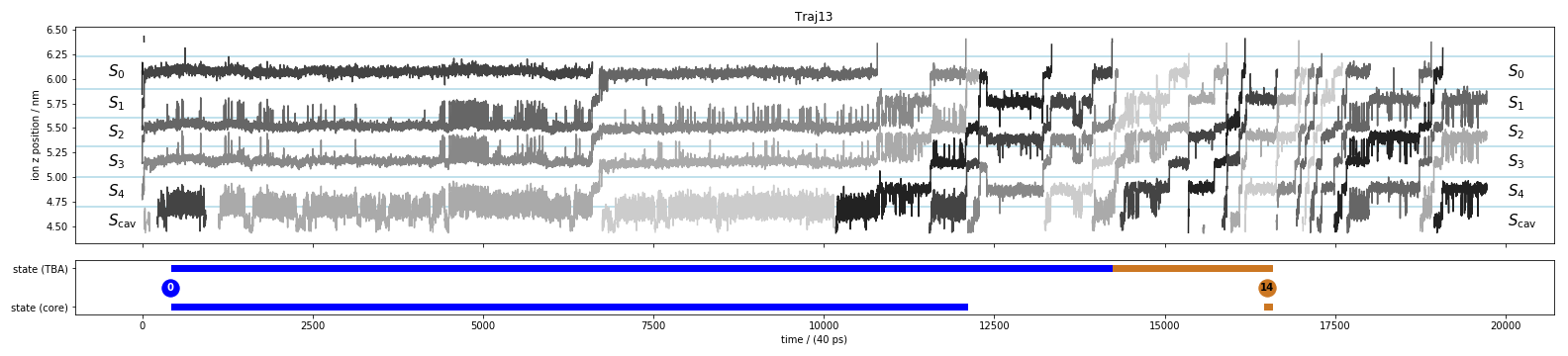}

\includegraphics[width=1\textwidth]{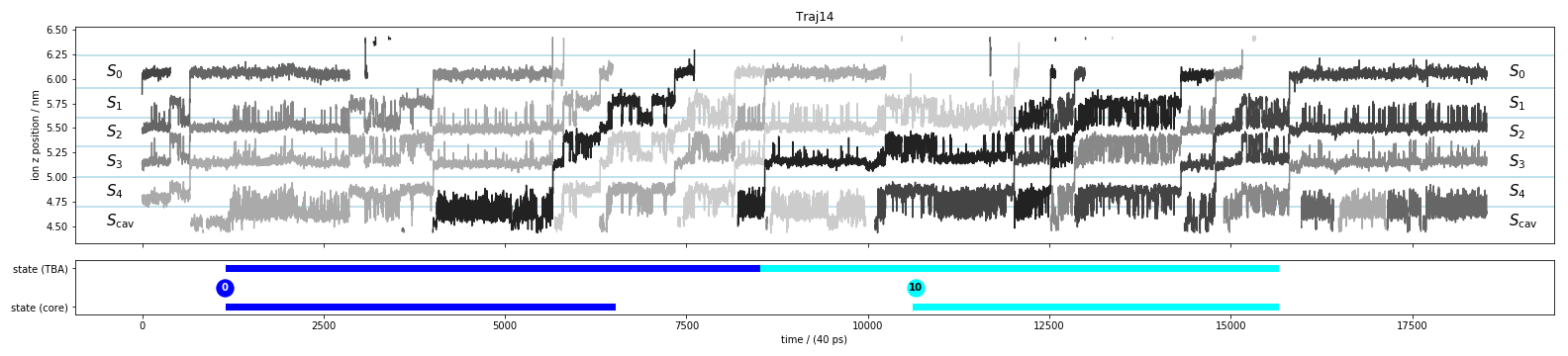}

\includegraphics[width=1\textwidth]{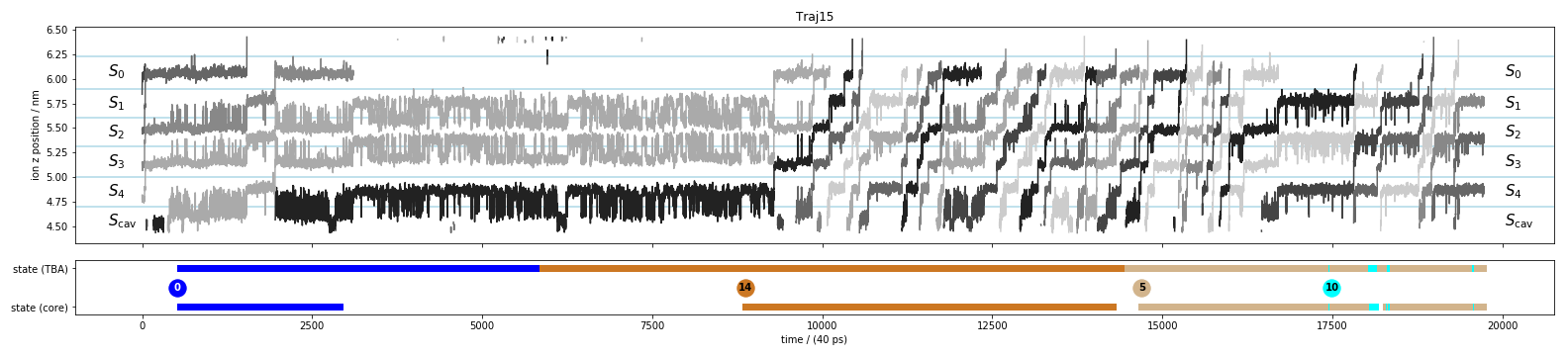}

\caption{\label{fig:time-traces-3}Ion positions and assignment to metastable
state for MD trajectories 11 to 15. For details see caption of figure
\ref{fig:time-traces-1}.}
\end{figure}

\begin{figure}[H]
\includegraphics[width=1\textwidth]{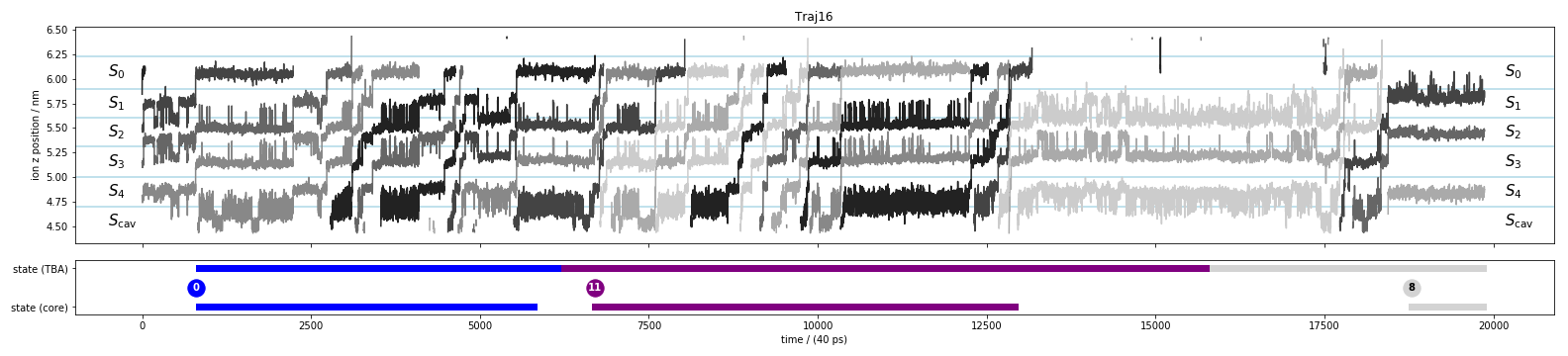}

\includegraphics[width=1\textwidth]{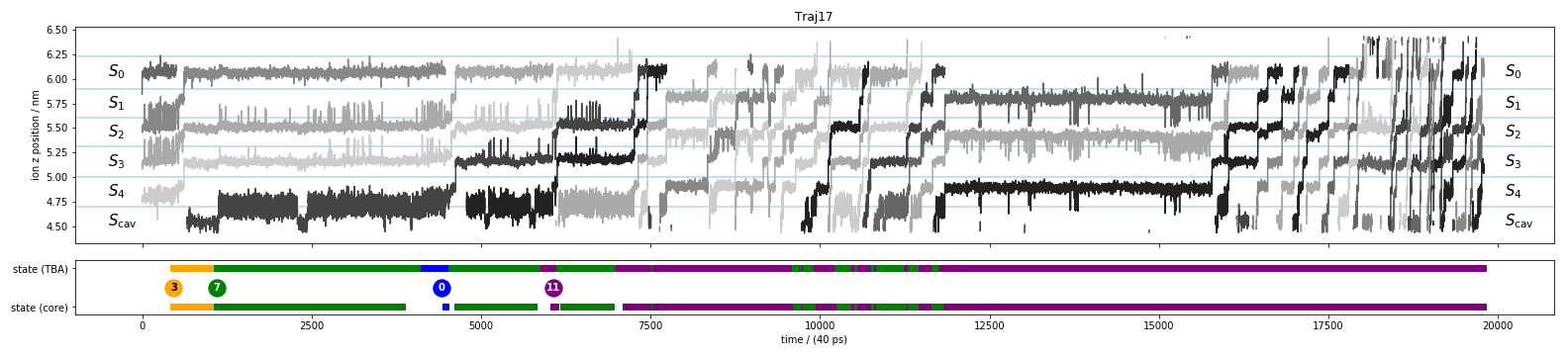}

\includegraphics[width=1\textwidth]{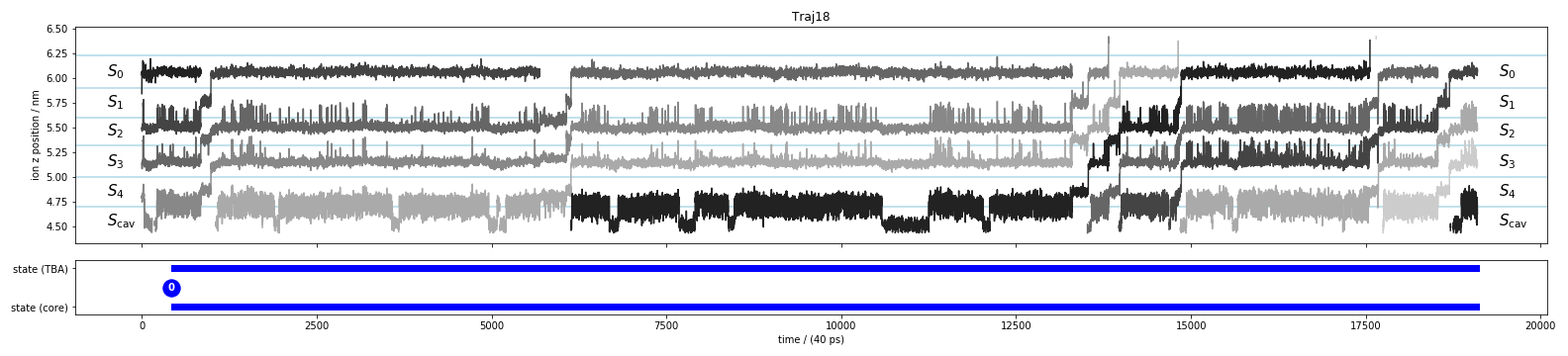}

\includegraphics[width=1\textwidth]{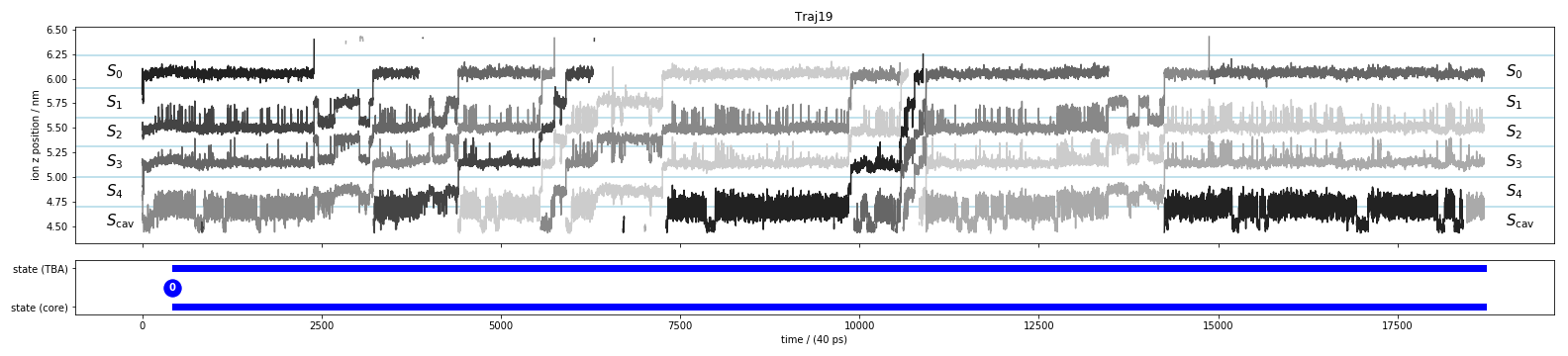}

\includegraphics[width=1\textwidth]{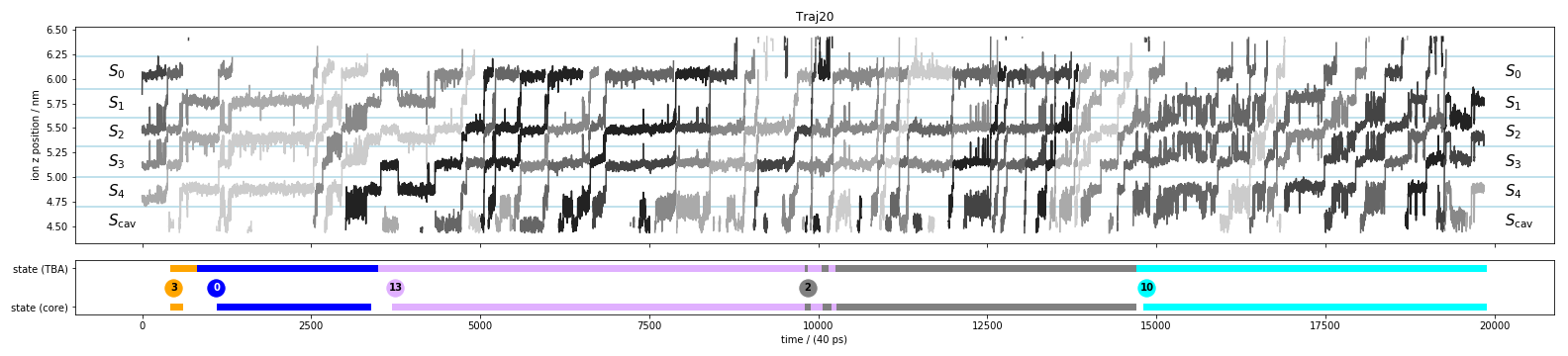}

\caption{\label{fig:time-traces-4}Ion positions and assignment to metastable
state for MD trajectories 16 to 20. For details see caption of figure
\ref{fig:time-traces-1}.}
\end{figure}

\begin{figure}[H]
\begin{centering}
\includegraphics[width=0.8\paperwidth]{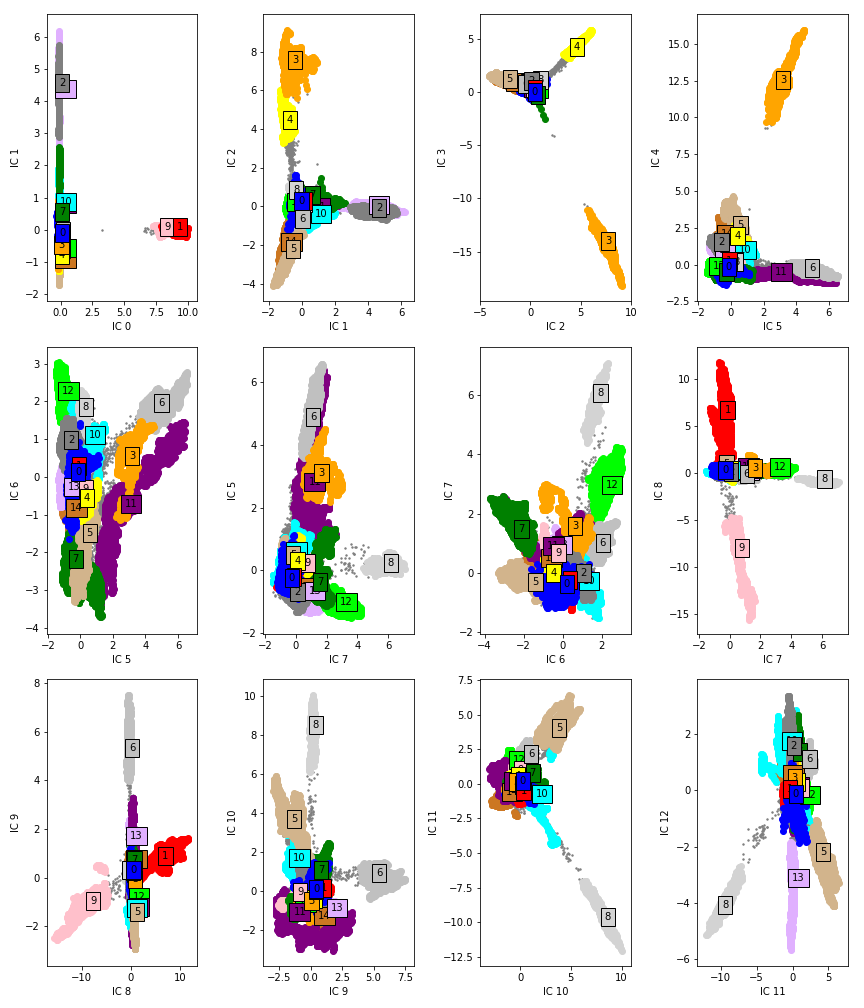}
\par\end{centering}
\caption{\label{fig:VAMP-projections-1}Projection of the simulation data on
pairs of singular functions (also called independent components) that
were computed with VAMP. Data points were colored according to the
metastable state to which they have the highest membership. Data points
that do not clearly belong to any of the metastable states (maximum
membership $<0.6$) are shown as small gray points.}
\end{figure}

\begin{figure}
\begin{centering}
\includegraphics[width=0.8\textwidth]{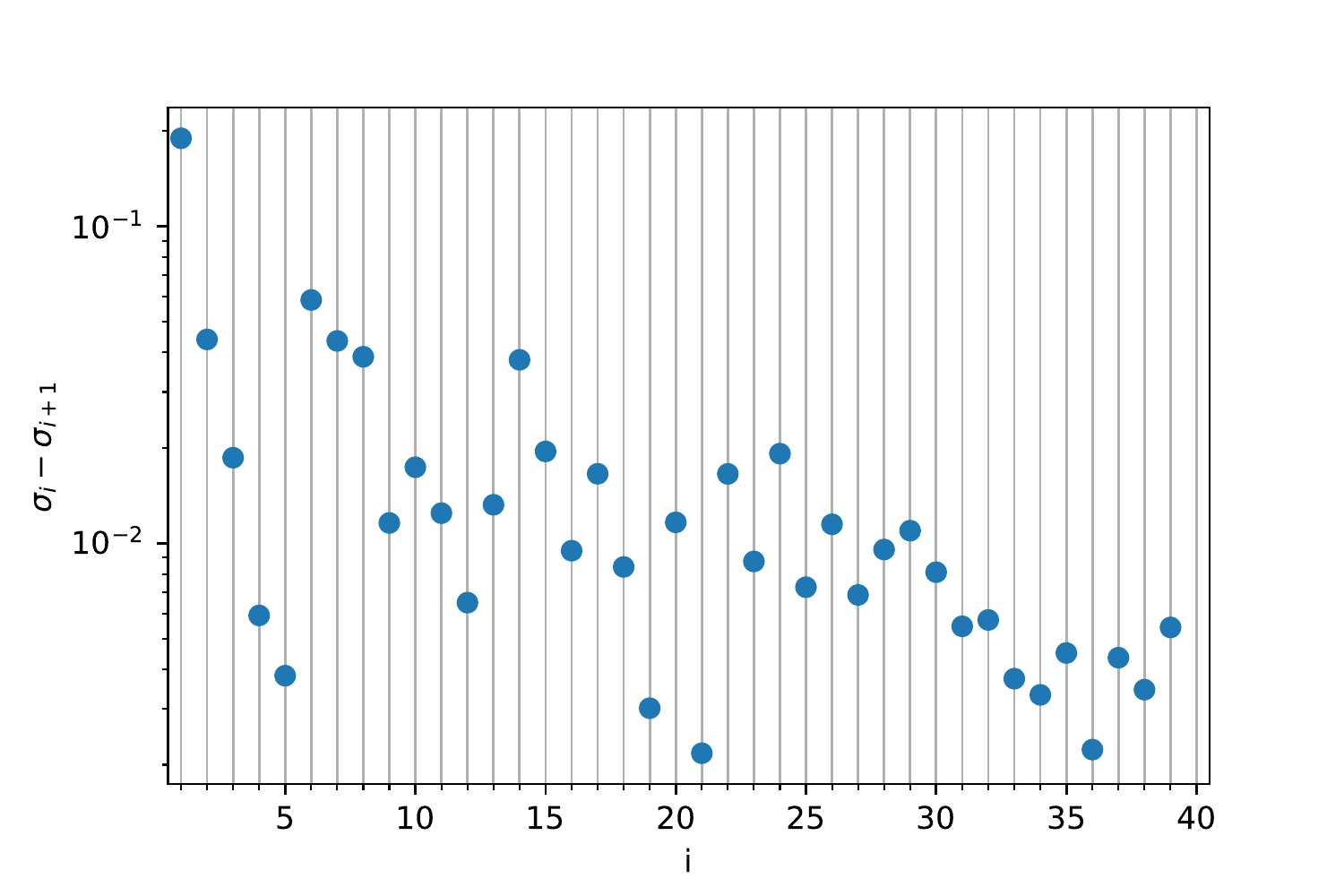}
\par\end{centering}
\caption{Jumps between successive generalized singular values of the Koopman
matrix for the KcsA channel. We observe larger jumps at positions,
1, 2, 6, 7 8 and 14. }
\end{figure}

\begin{figure}
\begin{centering}
\includegraphics[width=0.8\textwidth]{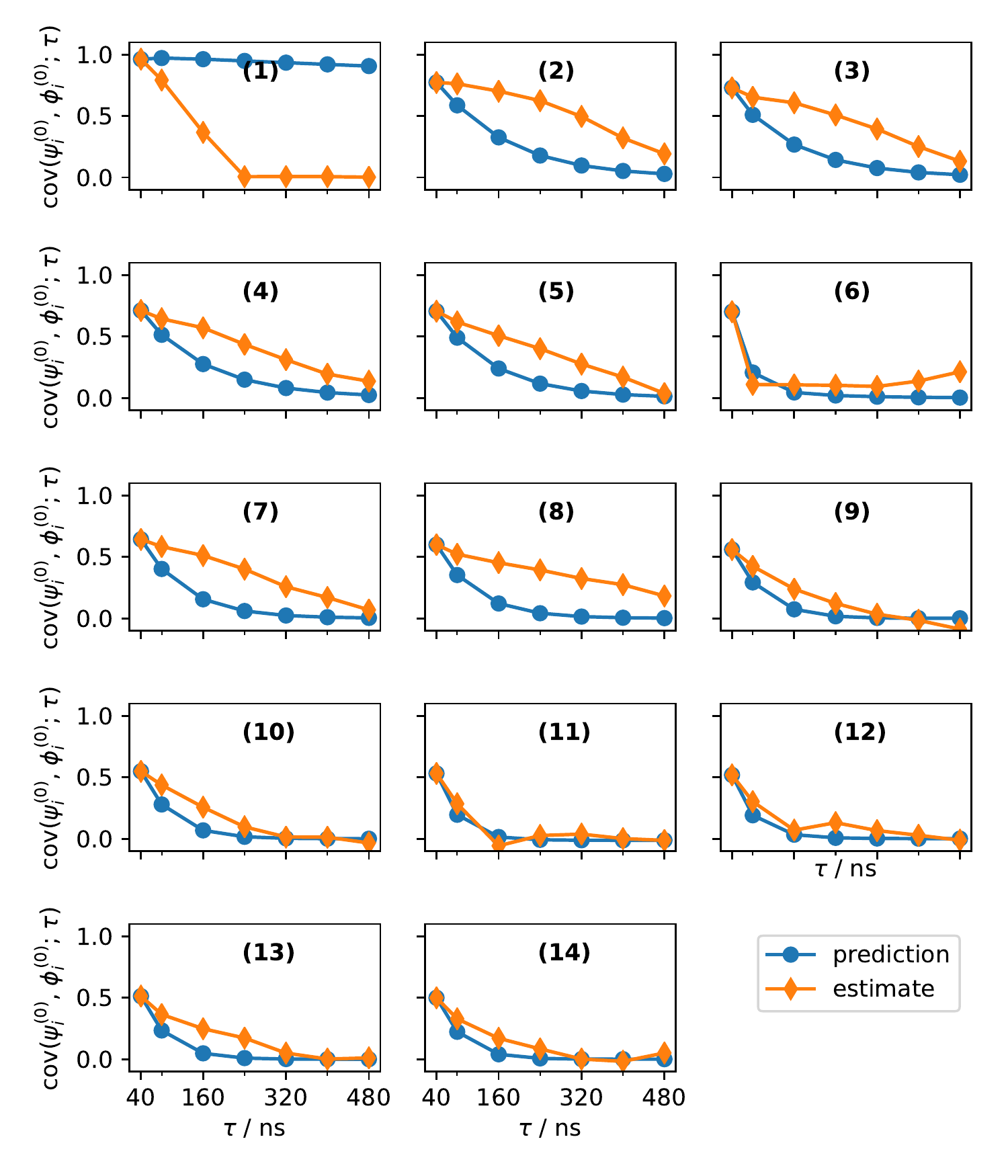}
\par\end{centering}
\caption{Non-equilibrium Chapman-Kolmogorov test for the dimensionality-reduced
Koopman model of KcsA filter conformational dynamics. The leading
left and right singular functions $\phi_{i}^{(0)}$, $\psi_{i}^{(0)}$
were computed at the lowest lag time $\tau_{0}=40\,\mathrm{ns}$.
For each pair, $(\phi_{i}^{(0)},\psi_{i}^{(0)})$, the time lagged-autocorrelation
is computed at integer multiples $n\tau_{0}$ in two ways: ``Estimates''
are computed by re-estimating the complete Koopman model from the
MD data at the new lag time and using it to compute the time-lagged
covariance of $\phi_{i}^{(0)}$ and $\psi_{i}^{(0)}$. ``Predictions''
of the time-lagged covariances are from the $n$'th power of the Koopman
matrix that was estimated at $\tau_{0}$. Bold numbers indicate the
index $i$ of the pair of singular functions.}
\end{figure}